\documentclass[a4paper]{article}
\usepackage{hyperref}
\usepackage{graphicx}
\usepackage{color}
\usepackage{qcircuit}
\usepackage{physics}
\usepackage{amsmath}
\usepackage{amssymb}
\usepackage{authblk}
\usepackage{subfig}
\usepackage{relsize}

\usepackage{amsmath}

\def\ket#1{|#1\rangle}

\newlength{\dhatheight}

\newcommand{\kket}[1]{|{#1}\rangle}

\usepackage{booktabs} 

\begin{document}
\title{Network architecture for a topological quantum computer in silicon}

\author[1,2,3]{Brandon Buonacorsi}
\affil[1]{Institute for Quantum Computing, University of Waterloo, Waterloo, Ontario, Canada N2L 3G1}
\affil[2]{Waterloo Institute for Nanotechnology, University of Waterloo, Waterloo, Ontario, Canada N2L 3G1}
\affil[3]{Department of Physics and Astronomy, University of Waterloo, Waterloo, Ontario, Canada N2L 3G1}
\author[4]{Zhenyu Cai}
\affil[4]{Department of Materials, University of Oxford, Parks Road, Oxford OX1 3PH, UK}
\author[1,2,3]{Eduardo B. Ramirez}
\author[1,2,3]{Kyle S. Willick}
\author[1,5]{Sean M. Walker}
\affil[5]{Department of Chemistry, University of Waterloo, Waterloo, Ontario, Canada N2L 3G1}
\author[1]{Jiahao Li}
\author[1]{Benjamin D. Shaw}
\author[4]{Xiaosi Xu}
\author[4]{Simon C. Benjamin}
\author[1,2,5]{Jonathan Baugh}

\maketitle
\date{\today}

\begin{abstract}
A design for a large-scale surface code quantum processor based on a node/network approach is introduced for semiconductor quantum dot spin qubits. The minimal node contains only seven quantum dots, and nodes are separated on the micron scale, creating useful space for wiring interconnects and integration of conventional transistor circuits. Entanglement is distributed between neighbouring nodes by loading spin singlets locally and then shuttling one member of the pair through a linear array of empty dots. A node contains one data qubit, two ancilla qubits, and additional dots to facilitate electron shuttling and measurement of the ancillas. A four-node GHZ state is realized by sharing three internode singlets followed by local gate operations and ancilla measurements. Further local operations produce an $X$ or $Z$ stabilizer on the four data qubits, which is the fundamental operation of the surface code. Electron shuttling is simulated in the single-valley case using a simple gate electrode geometry without explicit barrier gates, and demonstrates that adiabatic transport is possible on timescales that do not present a speed bottleneck to the processor. An important shuttling error in a clean system is uncontrolled phase rotation of the spin due to modulation of the electronic $g$-factor during transport, owing to the Stark effect. This error can be reduced by appropriate electrostatic tuning of the stationary electron's $g$-factor. While these simulations are unrealistic in neglecting spin-orbit, valley and decoherence effects, they are realistic with respect to the gate-induced potential landscape and are a first step towards more realistic modelling. Using reasonable noise models, we estimate error thresholds with respect to single and two-qubit gate fidelities as well as singlet dephasing errors during shuttling. A twirling protocol transforms the non-Pauli noise associated with exchange gate operations into Pauli noise, making it possible to use the Gottesman-Knill theorem to efficiently simulate large codes.
\end{abstract}

\section{Introduction}
Building a large-scale, universal quantum computer would enable major technological advances, yet presents a significant challenge. Solid-state qubits based on superconducting circuits \cite{gambetta2017building, wendin2017quantum}, semiconductor quantum dots \cite{hanson2007spins, kloeffel2013prospects}, semiconductor donor spins \cite{kane1998silicon, zwanenburg2013silicon, pla2012single}, or topologically protected quantum states \cite{sarma2015majorana} offer exciting prospects for a quantum computer chip, in analogy to classical CMOS devices. The standard circuit model for quantum computation, however, requires a staggering error correction overhead to achieve fault tolerance. Topological stabilizer codes acting on two-dimensional qubit arrays, i.e. surface codes \cite{raussendorffault2006, fujiiquantum2015}, can tolerate relatively high error thresholds and are considered one of the most promising approaches to scaling up. Fowler et al \cite{fowler2012surface} estimate that $\sim$ 100 million physical qubits would be required to factor a 2000 bit semiprime (i.e. RSA) number via Shor's algorithm on a surface code processor. In that estimate, the ratio of logical to physical qubits is $\sim 10^{-4}$. Scaling to this size, while maintaining the requisite precision of quantum control and the necessary cryogenic environment, is far beyond what is possible today. Superconducting qubit processors are rapidly advancing from several qubits \cite{neill2018blueprint, song201710, kandala2017hardware} to the 50-100 qubit scale, while competing platforms such as semiconductor quantum dots are still developing at the few-qubit scale \cite{watson2018programmable, veldhorst2015two, eng2015isotopically, yoneda2018quantum, zajac2016scalable}. Ultimately, the qubit footprint matters for a large-scale monolithic chip to be possible. Quantum dot and donor qubits have the advantage of a small (tens of nanometer) footprint compared to other platforms like superconducting or trapped ion qubits, making an area density of $\sim 10^{10}$ physical qubits per cm$^2$ a theoretical possibility. A rigorous analysis based on a compact exchange-only silicon double dot qubit, accounting for technological and physical constraints as a function of CMOS technology node, predicts that a $10^{10}$ cm$^{-2}$ density of physical qubits is possible at the 7 nm CMOS node, corresponding to $\sim 10^4-10^6$ cm$^{-2}$ logical qubits depending on the error correction scheme chosen \cite{rottaquantum2017}. The ability to integrate classical components on the quantum chip to facilitate multiplexing of control and readout signals will be advantageous. Semiconductor qubits also have an advantage in this respect, especially those based on silicon platforms. We will refer to realizing electron or hole spin qubits in a silicon MOS device structure \cite{maurand2016cmos, veldhorst2014addressable, harvey2018high, jock2018silicon} at cryogenic temperatures as `QMOS'. A QMOS approach can benefit from the vast investments and advances that have been made in conventional CMOS device processing, and is naturally compatible with CMOS integration. In this paper, we propose a QMOS architecture that is based on a network/node approach and is distinct from existing proposals \cite{hill2015surface, o2016silicon, pica2016surface, veldhorst2017silicon, li2017crossbar}. This approach is advantageous because it separates the surface code operation into two fundamental parts: local node operations that should be feasible to demonstrate in the near-term, and medium-range entanglement distribution that is more challenging but can be developed in parallel. Our scheme provides greater isolation of the data qubits than a conventional close-packed 2D array, and naturally opens up useful space to ease wiring density constraints and allow integration of supporting components to facilitate multiplexing of control and readout signals. \\
\indent While much early progress in quantum dot spin qubits was achieved in GaAs 2D electron gas devices, silicon offers the possibility of a nuclear spin free lattice, which has been demonstrated to yield electron spin coherence times of order seconds for donor electrons \cite{tyryshkin2012electron, muhonen2014storing} and up to tens of milliseconds for MOS quantum dot spin qubits \cite{veldhorst2015two}. The intrinsic spin-orbit interaction for electrons at the conduction band edge in silicon is weak compared to III-V semiconductors, which leads to longer spin relaxation and decoherence times. An enhanced spin-orbit interaction arises at the Si/SiO$_2$ interface due to inversion asymmetry leading to variation in the electronic $g$-factor, however this can be tuned near zero by the orientation of the external magnetic field \cite{jock2018silicon}. The variation in $g$, of order $10^{-2}$ at most \cite{veldhorst2015spin, ferdous2018interface, veldhorst2014addressable}, is tunable by the vertical electric field strength and can be used for addressing individual spins with a global microwave ESR field, or as a second control axis for singlet-triplet qubit rotations \cite{jock2018silicon}. Disadvantages of silicon compared to III-V's include the valley degeneracy problem \cite{culcer2010quantum, yang2013spin} and greater difficulty in accurately modelling two-qubit exchange energies \cite{li2010exchange}. Valley splittings are enhanced at interfaces, and have been observed for MOS dots up to several hundred $\mu$eV but vary considerably depending on local electric fields and disorder \cite{gamble2016valley, lim2011spin}. While Si/SiGe quantum wells present less disorder in the electrostatic potential and are thus `cleaner', valley splittings are found to be smaller on average for quantum dots in this material \cite{borselli2011measurement, neyens2018critical, zajac2015reconfigurable}. \\
\indent While for MOS quantum dots the microscopic roughness of the SiO$_2$ interface leads to an unavoidable degree of intrinsic variation in electrostatic and qubit parameters, the large scale uniformity of the Si/SiO$_2$ material system is remarkable and has been critically important to the scaling of classical CMOS. Many engineering challenges, however, can be foreseen with developing large scale QMOS: (i) qubit sensitivity to charge noise, (ii) control line cross-talk, (iii) variability in device tuning parameters, (iv) need for high density 3D wiring interconnects, (v) need for multiplexing and parallel operations, (vi) ultra-low power dissipation, (vii) high precision / high bandwidth / low noise voltage controls, etc. Existing proposals make use of 2D quantum dot arrays as a basis for a surface code quantum computer. Veldhorst et al \cite{veldhorst2017silicon} suggest a two-layer structure, with a closely-packed 2D dot array at a lower $^{28}$Si / SiO2 interface, and an upper Si transistor layer to enable a word-line/bit-line qubit addressing scheme using floating gates. Each dot is singly charged and has four nearest neighbours with exchange interactions that must be separately controlled. Single-qubit rotations are achieved via global microwave field and gate tuning of individual electronic $g$-factors. This approach utilizes shared control lines and is therefore scalable in principle, but requires a high interconnect density with feature sizes well below present technological capabilities. All qubits, both data and measure, experience the same local noise environment and capacitive cross-coupling to many electrodes, so that both control line cross-talk and gate voltage noise would present challenges. Furthermore, the power dissipated by conventional transistors would make it difficult to maintain milliKelvin temperatures, either requiring very large cooling powers or qubit operation at temperatures approaching 1 K. Li et al \cite{li2017crossbar} propose an alternate scheme using shared control that makes use of a half-filled 2D lattice, so that between operations qubits are better isolated. It relies on shuttling electrons between adjacent lattice sites to accomplish two-qubit interactions, and uses dc currents in a subset of control lines to tune local ESR frequencies in concert with a global ESR field. Since the dots and tunnel barriers are all controlled by a crossed array of common lines, this scheme requires a high degree of device uniformity, at least an order of magnitude beyond what has yet been demonstrated in experiments. To avoid practical issues with scaling a qubit array beyond $\sim 1000$ qubits, it was proposed to join arrays in a network, making use of electron shuttling `highways' consisting of linear dot arrays to transmit quantum information. Hence, both local and long-range electron shuttling are critical elements of Li et al's proposal. Our scheme also makes essential use of electron shuttling to distribute entanglement between adjacent nodes; however, these are small nodes of fewer than 10 quantum dots, so that the nodes and their corresponding local quantum operations are nearly within the reach of present experimental capabilities. \\

\begin{figure}[t!]
\centering
\includegraphics[width = 0.65\linewidth]{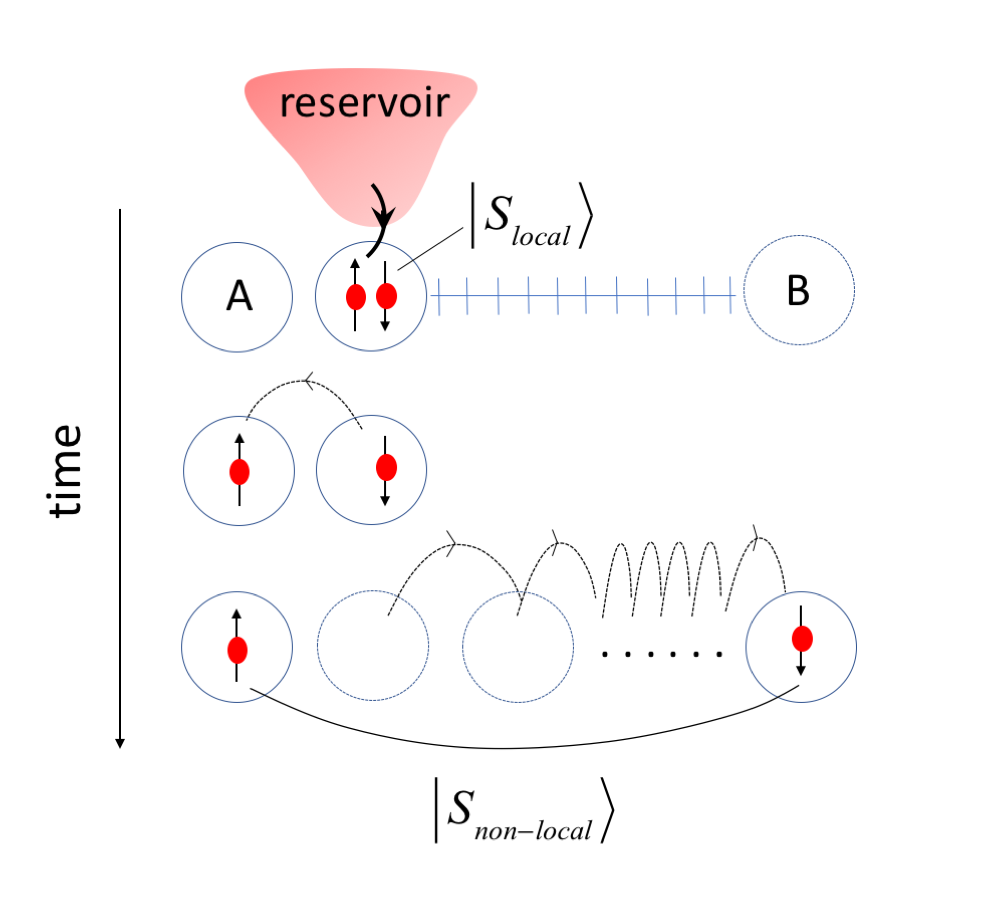}
\caption{\label{fig1} Spatial separation of the spin singlet state $\ket{S}$ across distant quantum dots A and B, via spin shuttling through a linear chain of normally empty quantum dots. The two-electron ground state singlet is loaded into a quantum dot tunnel coupled to the reservoir. The singlet is separated into a (1,1) charge state with one electron in dot A, then the other electron is shuttled to a distant dot B. Both the weakness of the spin-orbit interaction for conduction electrons in silicon and the isotopic removal of $^{29}$Si nuclear spins help to preserve spin coherence during transport over micron scales. }
\end{figure}

\begin{figure}[t!h!]
\centering
\includegraphics[width = \linewidth]{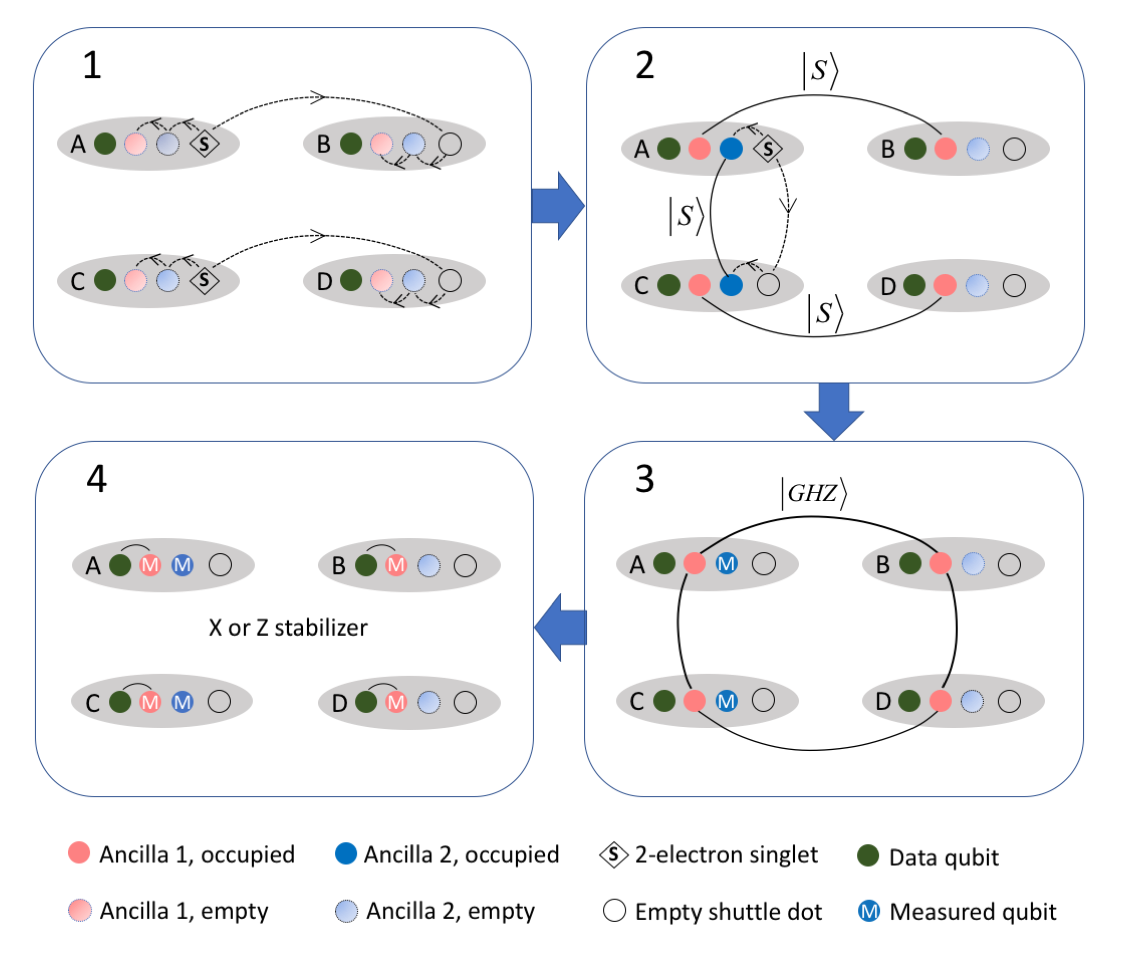}
\caption{\label{fig2} Sequence of steps in the stabilizer operation on four neighbouring nodes labeled A-D. Prior to step 1, all dots are empty except for the data qubit dots. In \textbf{step 1}, singlets created at nodes A and C are shared between nodes A/B and between C/D, populating the ancilla 1 qubits. Long dashed lines indicate internode shuttling. \textbf{Step 2}: a singlet created at A is shared between ancilla 2 qubits on nodes A/C. \textbf{Step 3}: ancilla 2 qubits on nodes A/C are measured, which projects the four ancilla 1 qubits into the shared GHZ state with probability 1/2, or with equal probability into a state that is transformed to the GHZ state by local gate operations. \textbf{Step 4}: conditional quantum gates (control-NOT or control-$Z$) are performed between ancilla 1 and data qubits, followed by measurement of the ancilla 1 qubits, realizing a stabilizer operation on the data qubits. Step numbers are color-coded to match circuit segments in figure 4.  }
\end{figure}

\begin{figure}[t!h!]
\centering
\includegraphics[width = 0.9\linewidth]{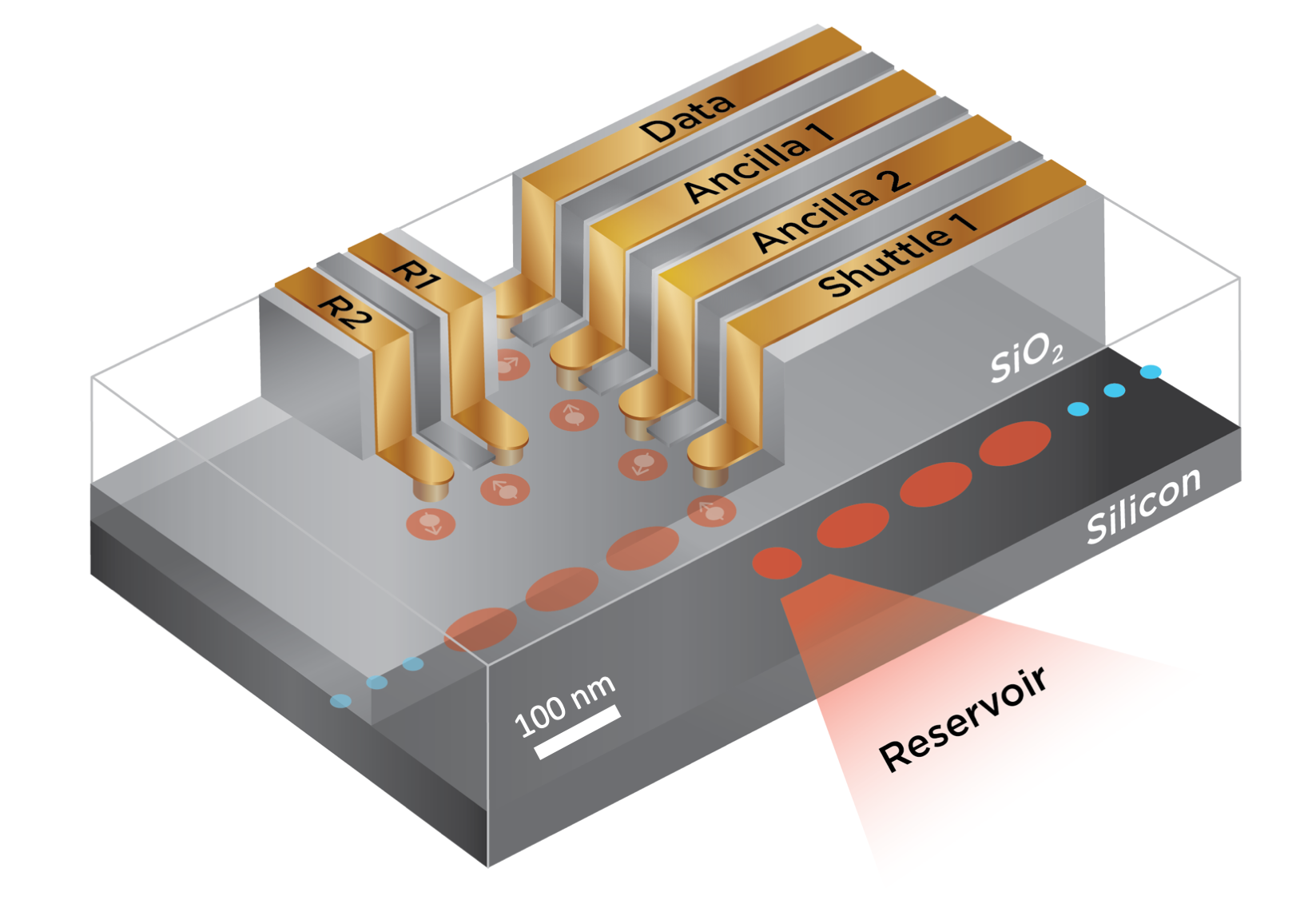}
\caption{\label{fig3} Device concept for a node. Accumulation mode MOS quantum dots are formed by single `via' gate electrodes (gold color), with additional gates to control interdot tunneling (silver color). All quantum dots (red circles and ovals) form in Si just below the interface with SiO$_2$. For clarity, gate electrodes forming the electron shuttling pathways (oval dots) and the electron reservoir are not shown. The intra-node dots are identified by the labels on the via gate electrodes. In this version of the node there are two shuttle dots, one tunnel coupled to the shuttle path going left, the other to the path going right and to the reservoir. This geometry ensures no more than three tunnel couplings per dot. The labels $R1, R2$ indicate the double dot that allows for readout of the ancilla qubits, using the singlet-triplet spin basis together with RF reflectometry. Gate electrodes to control exchange between the ancilla qubits and $R1$ are not shown but are implied. The oval-shaped dots making up the shuttle pathway are each defined by a single gate, with no additional barrier gates.  }
\end{figure}

\begin{figure}[t!h!]
\centering
\includegraphics[width = 1.0\linewidth]{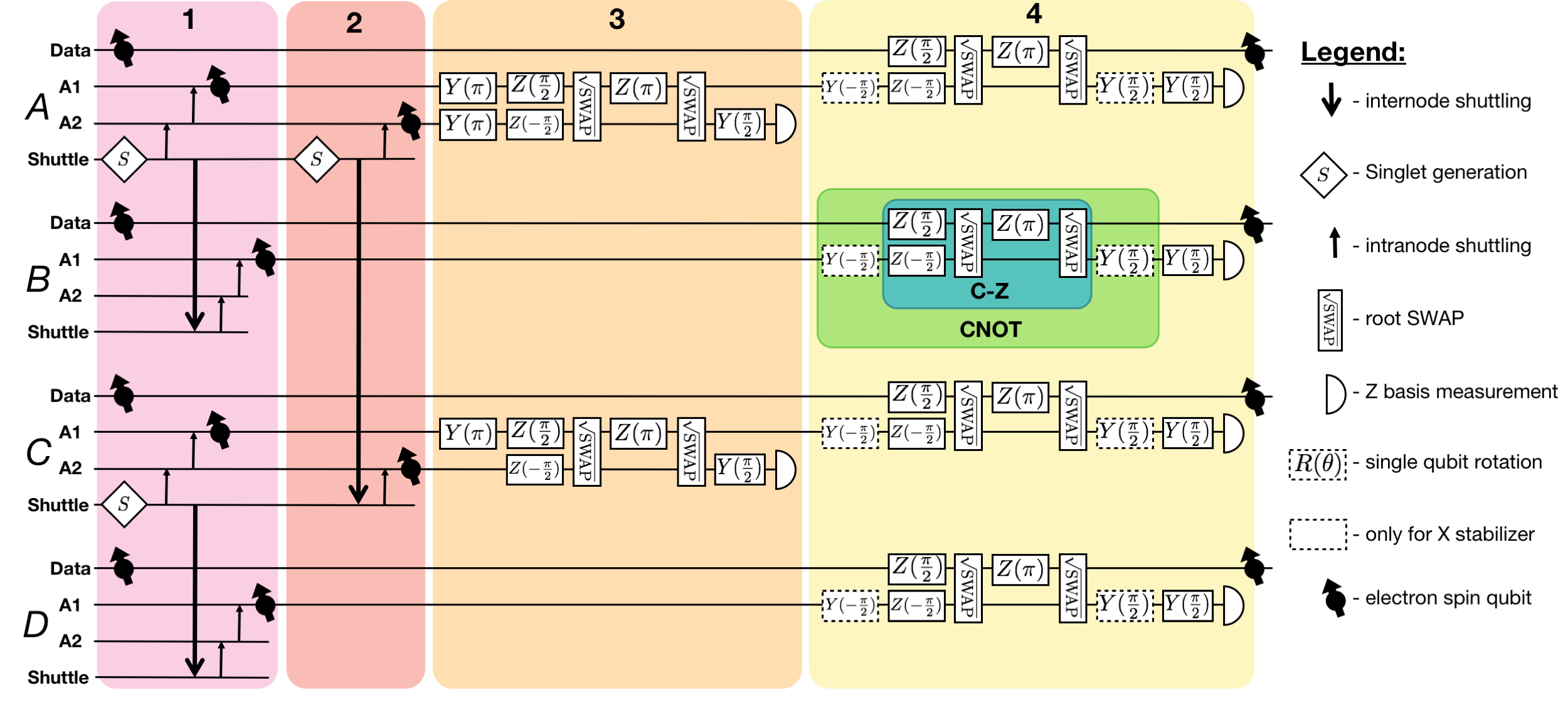}
\caption{\label{fig4} Circuit diagram corresponding to the four-node operations shown in figure 2. The labels 1-4 at the top correspond to the steps in figure 2; entanglement distribution is carried out in steps 1-2, the GHZ state is formed at the end of step 3, and the stabilizer operation is step 4. A1 and A2 refer to ancilla qubits 1 and 2, respectively. Other symbols are defined in the legend below. The notation $R(\theta)$ indicates a spin rotation about the $R$ axis in the Bloch sphere by an angle $\theta$. Control-$Z$ gates in step 4 correspond to a $Z$-stabilizer, whereas additional $Y(\pi/2)$ rotations transform these to control-NOT gates which yield the $X$-stabilizer. In the diagram we assume the ability to perform single qubit rotations on the data and both ancilla qubits, however, use of additional SWAP gates could restrict this requirement to one qubit, e.g. single qubit gates on A1 only. The control-$Z$ sequences could be replaced by direct gates under certain circumstances \cite{watson2018programmable}, reducing the number of single-qubit gates and increasing processor speed. Steps to empty the ancilla dots are not shown explicitly, but will directly follow the final measurements.}
\end{figure}

\section{Node/network surface code for quantum dots}
It is well understood that a universal quantum computer could be constructed by networking together many simple processor cells, rather than building a single complex device \cite{nickerson2013topological, kimble2008quantum, cirac1999distributed, nickerson2014freely}. Key to this approach is the ability to distribute entanglement between such cells, or network nodes. Nickerson et al \cite{nickerson2013topological} showed that even with realistically noisy entanglement distribution, with raw error rates approaching 10$\%$, entanglement purification strategies could be used to reduce the effective error rates to tolerable levels. Combined with sufficiently high fidelity local gate operations, state preparation and measurement, a stabilizer protocol was described that enables a two-dimensional surface code to be implemented \cite{nickerson2013topological}. This method is straightforwardly applicable to systems like trapped ion qubits, where spatially separated traps can be linked photonically \cite{hucul2015modular, brown2016co, monroe2014large}. Successful entanglement distribution via photonic link is currently probabilistic and slow, however, with typical rates on the scale of a few Hz, limiting practical processor speeds. Here we propose to apply the network model to a monolithic silicon QMOS chip, with internode distance on the micron scale. We exploit the natural property of spin qubits to form a singlet ground state in a doubly occupied quantum dot to create the entanglement resource, and the weak spin-orbit interaction in silicon to allow coherent shuttling of electron spins via interdot tunnelling, as illustrated in figure~\ref{fig1}. Thus, entanglement distribution becomes effectively deterministic. Although our approach is monolithic and thus returns to `building a single complex device', we gain significant advantage by separating the scaling problem into two distinct parts, and by creating useful space between these very compact qubits to improve qubit isolation and make wiring/integration more practical. Numerical simulations show that electron shuttling on the micron scale can be carried out with high fidelity in principle, and on the timescale of single-qubit ESR gate operations so that shuttling does not create a speed bottleneck. Further, we show that phase error in the singlet state due to Stark effect modulation of the $g$-factor during shuttling can be reduced to negligible levels with appropriate electrostatic tuning. Finally, we obtain threshold values for errors in gate and shuttling operations that would be required for a scaled up network to be fault tolerant, using reasonable noise models and the Gottesman-Knill theorem \cite{gottesman1998heisenberg, aaronson2004improved} to efficiently simulate large networks.  \\
\indent For simplicity, we will assume that spatial separation of the singlet states can be done with high fidelity, so that entanglement purification is not needed. This allows for a minimal node consisting of one data and two ancilla qubits. Additional ancillae and entanglement distribution operations could be used for entanglement purification if needed, as described in ref.~\cite{nickerson2013topological}. A four-qubit GHZ state is formed across four neighbouring nodes, making use of singlet separation and the ancilla qubits in each node, as shown in figure~\ref{fig2}. The GHZ resource shared among ancilla 1 qubits, together with conditional logic gates (control-NOT or control-$Z$) applied to the data qubits, allows for the X or Z stabilizer operation to be carried out. In addition to the three quantum dots hosting the data and ancilla qubits, there are 1-2 additional dots to facilitate the distribution of singlet states, which we will refer to as `shuttle' dots. A conceptual device-level illustration of a node is shown in figure~\ref{fig3}. In this version, there are two shuttle dots, which ensures no more than three tunnel couplings per dot. The node is connected to a single electron reservoir via one of the shuttle dots, providing a means for initializing the charge state of the device and loading singlets into the shuttle dot prior to their distribution. All reservoirs are kept at a fixed potential of 0 V. The node in figure~\ref{fig3} is based on a simplified gate geometry in which each quantum dot is defined by a single `via' accumulation gate electrode. Additional barrier gates between intra-node dots allow for fine control of exchange. A double dot, aligned perpendicular to the data/ancilla linear array, allows for readout of both ancilla qubits, as will be described in section~\ref{sect:readout}. A global microwave field acts in concert with electrostatic tuning of the electronic $g$-factors to realize arbitrary single-qubit rotations via ESR. Dots forming the shuttle pathway are each formed by single gate electrodes, with no additional barrier gates, as we show in section~\ref{sect: shuttling}. \\
\subsection{Stabilizer circuit}
\label{sect:stabilizer}
\indent The four-node stabilizer sequence shown in figure~\ref{fig2} begins with all dots empty except for the data qubit. The circuit diagram corresponding to the stabilizer sequence is shown in figure~\ref{fig4}. A two-electron spin singlet state is loaded from the reservoir into the shuttle dots in nodes A and C, and then distributed across A-B and C-D via internode shuttling. This populates the ancilla 1 qubits. Next, a fresh singlet loaded in node A is distributed across the ancilla 2 qubits in nodes A-C. Control-NOT operations between ancilla 1 and 2 qubits on nodes A and C are carried out by a combination of single-qubit rotations and two-qubit exchange gates, i.e. $\sqrt{SWAP}$ gates. Subsequent $Z$-basis measurement of ancilla 2 qubits on nodes A and C projects the four ancilla 1 qubits into a maximally entangled GHZ state with probability 50$\%$, i.e. if the A and C measurements return even parity. If the parity is odd, then Pauli $X$ gates applied to the ancilla 1 qubits of nodes A and B will produce the GHZ state (see Appendix~\ref{sect:GHZ} for mathematical details). Thus, the protocol is deterministic subject to this feedback. The GHZ state provides the shared entanglement resource that allows the data qubits to be stabilized. A control-NOT (or control-$Z$) between the local ancilla 1 and data qubits, followed by measurement of the ancilla 1 qubits, performs a 4-qubit $X$ (or $Z$) stabilizer on the data qubits. The ancilla dots can then be emptied of electrons (via shuttling to the reservoir) to prepare for the next stabilizer operation. A full surface code cycle requires 4 separate stabilizer operations in sequence, since any two neighbouring 4-node plaquettes cannot be stabilized simultaneously. Both the $Z$ and $X$ plaquettes are split into two non-adjoining subsets, and each of the 4 subsets are stabilized sequentially (see Appendix~\ref{sect: scheduling}). As pointed out in ref~\cite{nickerson2013topological}, the stabilizer superoperator allows projectors and errors to be commuted so that errors occurring in between subsets can be corrected. Note that the control-NOT and control-$Z$ operations in steps 3 and 4 of figure~\ref{fig4} require single-qubit rotations on the data, ancilla 1 and ancilla 2 qubits. To simplify the device, however, one could restrict single-qubit control to ancilla 1 only, and use SWAP operations to realize gates on the neighbouring qubits. While this approach is more costly in terms of two-qubit gate error, it saves time since exchange gates are typically much faster than ESR rotations.

\begin{figure}[t!h!]
\centering
\includegraphics[width = 0.7\linewidth]{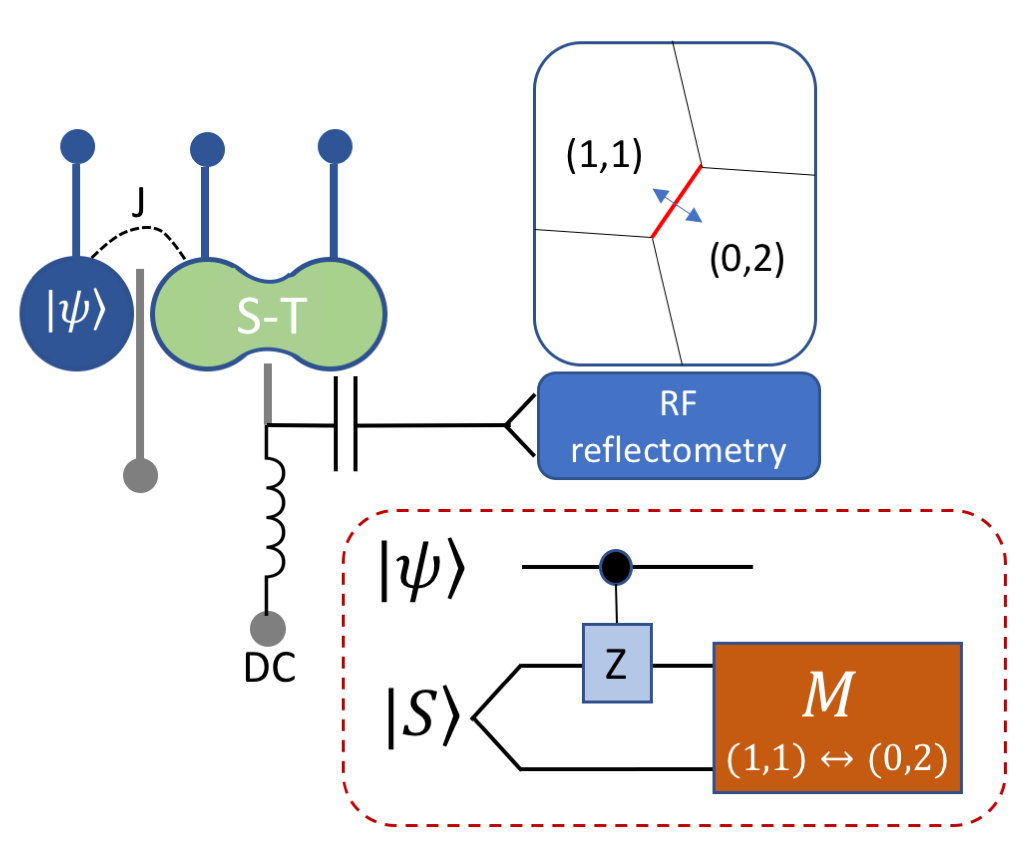}
\caption{\label{fig5} A method for reading out the ancilla qubits. A double quantum dot is operated in the two electron singlet-triplet basis (labeled S-T). A local gate electrode (gray) couples the quantum capacitance of the double dot to an RF reflectometry circuit for gate dispersive charge readout. A second (gray) gate electrode controls exchange (J) between the ancilla dot and the double dot. Initialized in the (1,1) singlet state, a control-$Z$ gate conditioned on the state $\kket{\psi}$ of the ancilla qubit acts on the dot adjacent to it. The ancilla states $\kket{0}$ and $\kket{1}$ thus map to the singlet and triplet states, respectively. The Pauli blockade prevents the T(1,1) state from tunnelling to the S(0,2) state, and thus the dispersive charge detection allows the two states to be distinguished. The conventional control-$Z$ sequence requires control of the exchange coupling and single-qubit rotations, the latter of which can be restricted to the ancilla qubit using SWAP gates.}
\end{figure}

\subsection{Readout of the ancilla qubits}
\label{sect:readout}
Measuring ancilla qubits quickly and with high fidelity is a critical requirement for any surface code processor, including the network approach proposed here. One method for projectively measuring the electron spin is to use spin-dependent tunneling together with a local charge sensor \cite{elzerman2004single, morello2010single}. However, this would require bringing both an electron reservoir and a charge sensor in close proximity to the ancilla qubits, both of which we aim to avoid in order to keep the data and ancilla qubits well isolated and reduce the number of local gate electrodes. Instead, we propose to use a double quantum dot placed so that it can be controllably tunnel coupled to both ancilla dots. The double dot is not coupled to a reservoir, but is coupled via local gate to an RF reflectometry circuit, as shown in figure~\ref{fig5}. The double dot is operated in the two-electron singlet/triplet basis. The readout sequence for the ancilla state $\kket{\psi}$ is the following: (1) initialize the double dot in the singlet (0,2) charge configuration, (2) separate into the (1,1) singlet, (3) perform a control-$Z$ gate operation between the ancilla and the adjacent member of the double dot, (4) tune the double dot to favour the (0,2) configuration and use gate-dispersive RF reflectometry \cite{petersson2010charge, rossi2017dispersive, gonzalez2016gate} to distinguish the T(1,1) and S(0,2) spin(charge) states. This charge detection method works by sensing the quantum capacitance associated with interdot tunnelling. When the ancilla qubit is in the logical $\kket{1}$ state, the control-$Z$ gate rotates the singlet to a triplet, which remains in the (1,1) charge state due to the Pauli spin blockade. The conventional control-$Z$ gate sequence requires two-qubit exchange and single-qubit rotations on both qubits, but it may be advantageous to restrict single-qubit rotations to the ancilla qubit by using SWAP gates. We note that single electron charge detection using gate-dispersive methods has demonstrated sensitivities allowing for measurement on few-nanosecond timescales \cite{gonzalez2015probing}, therefore, qubit readout times could be limited by the gate operations and not by charge detection. The presence of valley states in silicon complicates the spin-blockade based readout but is not a fundamental obstacle to achieving high readout fidelities \cite{tagliaferriimpact2018}.\\
\begin{figure}[t!h!]
\centering
\includegraphics[width = 1.0\linewidth]{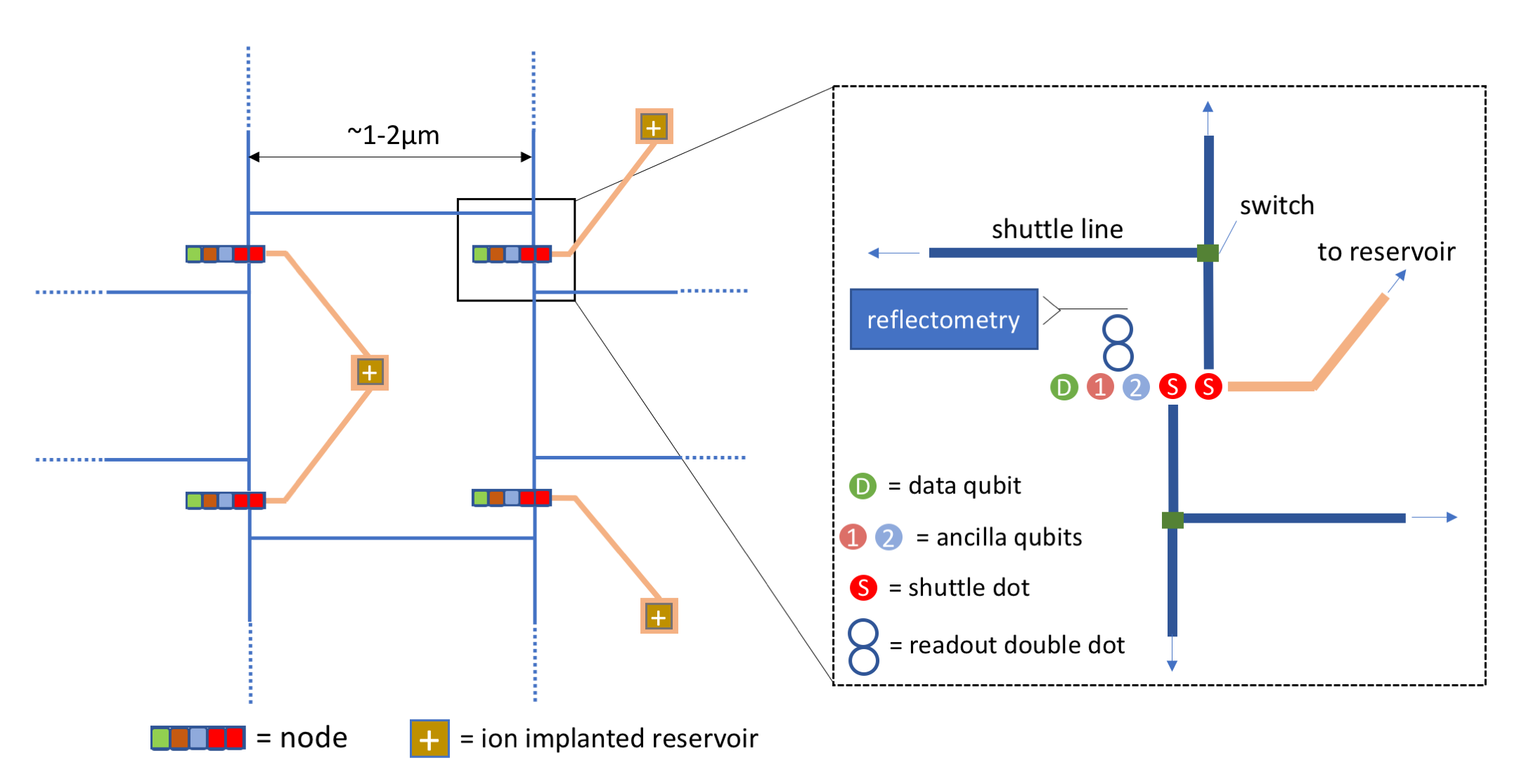}
\caption{\label{fig6} Illustration of a proposed network layout showing a plaquette of four nodes and how they connect beyond to form a 2D surface code. Ion implanted regions indicated by `+' symbols provide electron reservoirs that are brought to each node with accumulation gates (gold color). The enlarged section at right shows the dot layout in each node relative to the reservoir and shuttling paths, the latter here indicated as lines that represent linear (empty) dot arrays. Each switch is a T-junction of quantum dots in which the tunnelling direction is controlled by local gate electrodes. Gate electrodes that form the dots and enable quantum control (not shown) connect vertically (out of plane) to wiring in upper interconnect layers. Gate electrodes controlling shuttling can be shared, since singlet distribution occurs in parallel across the entire device. The RF reflectometry circuit indicated by the blue box represents a combination of on-chip and off-chip components and probes the charge state of the double dot by the gate-dispersive readout technique. Nodes in the main figure are not to scale. }
\end{figure}
\subsection{Network layout}
A proposed layout of the nodes forming a network is presented in figure~\ref{fig6}. N-type ion implanted regions, kept well separated from the nodes to reduce charge noise, allow reservoirs to be brought to each node using accumulation gates. The shuttle dots in each node connect to north/south shuttle pathways (linear dot arrays). The version shown here and in figure~\ref{fig3} has two shuttle dots so that no dot has more than three tunnel couplings that must be separately controlled. The data qubit dot is coupled only to the first ancilla, providing isolation for this all-important qubit. East/west shuttle paths can be chosen at T-junctions, where local gate electrodes control the tunnelling direction. Thus, each node is connected to all four neighbouring nodes. The internode distance can be scaled to optimize wiring density and integration of classical CMOS components while minimizing shuttle errors, and we expect this to be on the scale of $\sim 1$ to a few microns. For an internode spacing of 1.5 $\mu$m, the node (data qubit) density is 4.4$\times 10^7$ cm$^{-2}$, still a high density compared to superconducting and ion trap qubits. It is about 2 orders of magnitude less dense than the estimates given in Ref.~\cite{rottaquantum2017} for close-packed qubits, but would still give a few times $10^3$ cm$^{-2}$ logical qubits, enough to factor a 2000-bit semiprime number using ShorÕs algorithm \cite{rottaquantum2017, fowler2012surface}. The internode space could be used to add floating gate circuits to correct for small electrostatic variations in qubit device parameters, allowing for widely shared control lines. \\ 
\begin{figure}[t!h!]
\centering
\includegraphics[width = \linewidth]{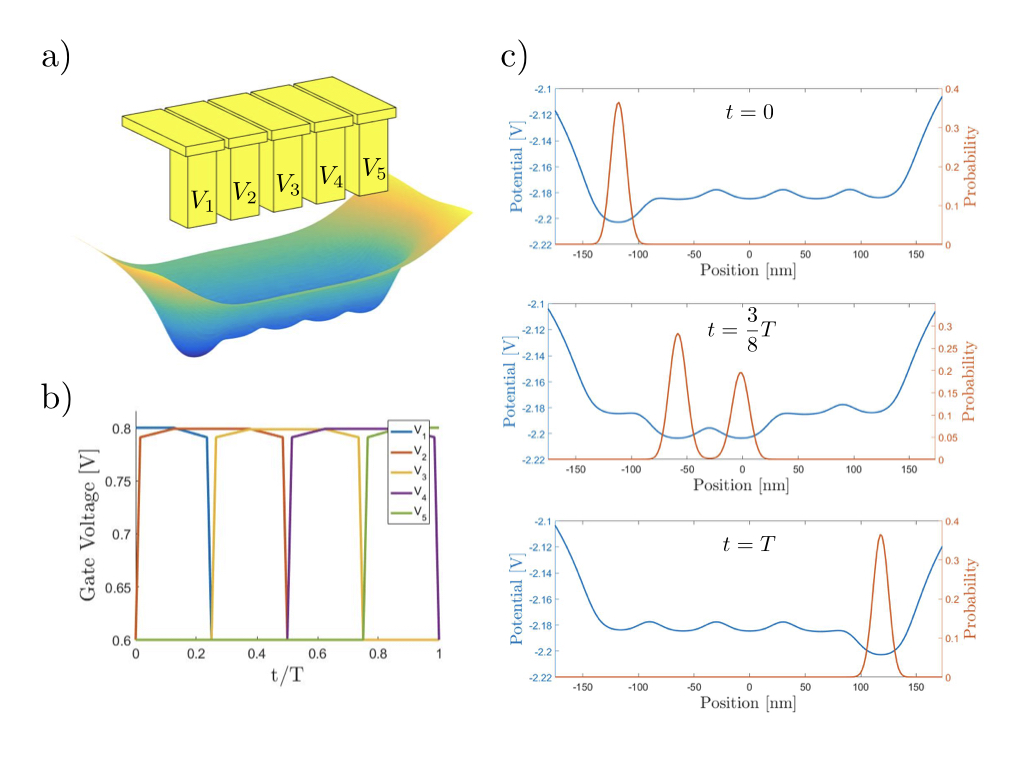}
\caption{\label{fig7} Setup for the electron shuttling simulations. (a) A linear chain of five quantum dots is formed by single `via' gate electrodes with applied voltages $V_1$-$V_5$. The corresponding electrostatic potential in Si just below the Si/SiO$_2$ interface is shown. (b) Gate voltage sweep sequence for moving an electron from dot 1 (left) to dot 5 (right), with time $T/4$ per shuttle. Voltage is swept more slowly near the resonant tunnelling point to preserve adiabaticity. (c) Snapshots of the 1D potential and ground state electronic probability density $|\Psi|^2$ at $t = 0$, $3T/8 $ and $T$. Note that potential differences are meaningful but the potential offset (absolute value) is arbitrary, and the sign of the potential is reversed as though the electron charge were positive. Coherent tunnelling between dots 2 and 3 can be seen at $t = 3T/8$. The minimum $T$ for which shuttling remains adiabatic is determined by the size of the tunnel coupling, the orbital energy spacing and other factors. }
\end{figure}
\section{Single electron transport}
\label{sect: shuttling}
Spatial separation of the spin singlet pairs is fundamental to the proposed network approach, and occurs in parallel across the device at the beginning of every stabilizer cycle. To coherently translate an electron spin across a distance requires confinement of the wavefunction be maintained. Single electron transport via `moving quantum dots' has been realized with surface acoustic waves in piezoelectric materials \cite{kataoka2009coherent, utko2007single, bertrand2016injection}. This idea was recently applied even in silicon, with an appropriate piezoelectric material attached to the surface \cite{buyukkose2013ultrahigh}. Surface acoustic wave generation, however, requires bulky interdigitated electrodes, and confining the waves to desired pathways is challenging. A more exotic possibility is the generation of a soliton wave \cite{keeling2006minimal, dubois2013minimal}, which would render unnecessary the requirement for a moving potential well. Unfortunately, solitons can only be generated from a Fermi sea, and not (as far as we know) from single particle levels in quantum dots. To create a moving confining potential without acoustic waves, one can use a set of gate electrodes to form a linear array of quantum dots \cite{fujita2017coherent, flentje2017coherent}. In the limit of many fine gate electrodes, a moving dot could be approximated. With realistic gate dimensions, however, it is more practical to define adjacent dots and force electrons to tunnel successively between them. We adopt a simplified model in which each dot is formed by a single accumulation (plunger) gate, and there are no explicit gates to control tunnelling. Instead, plunger gate voltages and the electrode geometry are used to control tunnelling. Two main topics are addressed: (1) what shuttling speeds are feasible in realistic devices while adiabatically maintaining the electronic ground state, and (2) how large is the shuttle-induced modulation of the electronic $g$-factor due to the Stark effect, how much error does this cause in the singlet state fidelity, and can it be mitigated? Although unrealistic for silicon \cite{tagliaferriimpact2018}, we assume a single valley model in this paper as a first step. Coherent spin transport through a series of dots is unlikely to succeed in cases for which the energy splitting between the two lowest valley states, $\Delta E_{vs}$, is comparable to the Zeeman and/or tunnelling energies, $E_z$ and $\epsilon_t$, respectively. In such cases, even a weak spin-orbit coupling causes levels with different spin and valley indices to anti-cross, so that diabatic transitions that mix spin and valley states are difficult to avoid \cite{zhaocoherent2018, liintrinsic2017}. Thus, our approach would require that $\Delta E_{vs} >> E_z, \epsilon_t$, so that the higher valley state would play a role similar to an excited dot orbital state. This condition is more likely to be achievable in MOS dots compared to Si/SiGe. The $g$-factor modulation is an indirect effect of the spin-orbit coupling and causes a phase rotation of the singlet state. Direct spin-orbit induced rotations along $\hat{x}, \hat{y}$ are expected to yield weaker errors, but are non-negligible for long shuttle paths, as we discuss below. Charge decoherence is also neglected in our simulations, but will be an important factor to consider in future work. As a side note, there is a closely related method referred to as coherent transfer by adiabatic passage (CTAP) which is analogous to the STIRAP technique in optics for population transfer in a three-level $\Lambda$ system \cite{greentreecoherent2004, rahmanatomistic2009}. CTAP, in a 3-dot linear array, relies on quantum interference to transfer an electron from dot 1 to dot 3 without it ever being in dot 2. This can be generalized to an N-dot system (for odd N). This method, however, is not feasible with the simplified gate geometry of our simulations because CTAP requires independent control of tunnel couplings and dot potentials, implying more gate electrodes are needed. CTAP is also sensitive to dephasing throughout the entire sequence, whereas shuttling is only sensitive during the tunnel events. For these reasons we have not included CTAP in our simulations, but it remains a possible alternative. \\
\begin{figure}[t!h!]
\centering
\includegraphics[width = 0.9\linewidth]{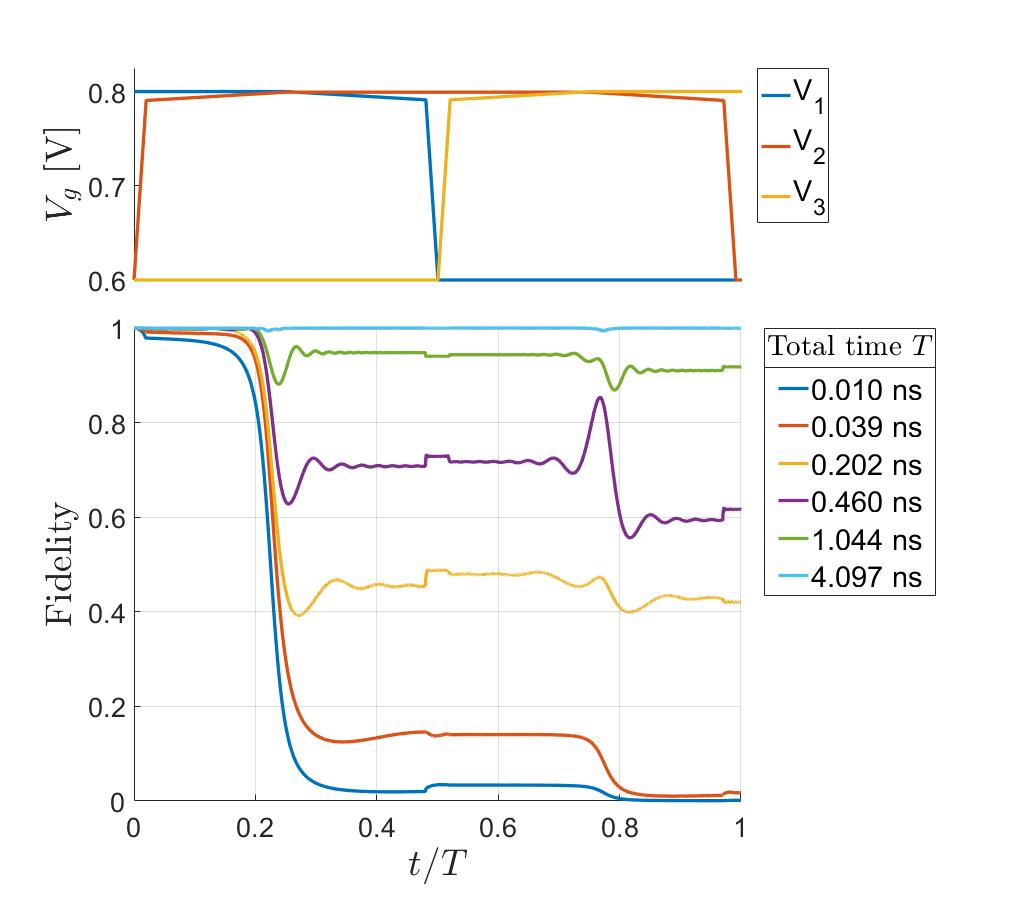}
\caption{\label{fig8} Adiabaticity threshold for a 3-dot shuttling simulation (i.e. two shuttle steps). The fidelity between the ground state wavefunction $\Psi_g(t)$ and the actual wavefunction $\Psi(t)$ is defined as $F = |\langle \Psi_g(t) | \Psi(t)\rangle|^2$. The total sequence time is $T$, and the normalized time $t/T$ is given on the horizontal axis. The corresponding gate voltages are shown in the panel above, sharing the same time axis. The process is adiabatic for $T = 4.1$ ns, but non-adiabatic for $T \leq 1$ ns. For a final state fidelity $F > 0.99$, the threshold for these simulations is $T_{th} \approx 3$ ns, or roughly 1.5 ns per shuttle step. }
\end{figure}
\subsection{Shuttling simulations}
\indent Figure~\ref{fig7}(a) shows an example of the gate electrode geometry and potential landscape for a five-dot linear array. This is simulated using a 3D self-consistent Poisson equation solver in the Nextnano software \cite{birner2007nextnano}. The `via' gate electrodes are 40 nm wide at the base, with center-center separation of 60 nm. The base of the via gate is separated from the Si interface by 17 nm of SiO$_2$, and the potential profile is shown 0.5 nm below the Si/SiO$_2$ interface. Figure~\ref{fig7}(b) shows the sequence of gate voltages applied in a shuttling simulation, with $V_1$ ($V_5$) corresponding to the leftmost (rightmost) gate electrodes. Resonant tunnelling from dot 1 to dot 2 occurs when $V_2 \simeq V_1$; voltages are swept slowly near this zero detuning point so that the electron remains in the ground state of the double dot. The same sequence then repeats for moving the electron from dot 2 to 3, etc, with small adjustments to take into account cross-capacitance effects (see section~\ref{sect: finetune} in Appendix). \\
\indent Figure~\ref{fig7}(c) shows the central one-dimensional (1D) slice of the electrostatic potential that was used for shuttling simulations, together with the electronic ground state wavefunction at three different points in the sequence (i.e. the actual wavefunction in the ideal adiabatic limit $T\rightarrow \infty$). The snapshot at $t = 3T/8 $ shows the electron tunnelling between dots 2 and 3. Here, the tunnel coupling is $\epsilon_t = 25~\mu$eV, giving a resonant tunnel rate $\Gamma = 24$ GHz, based on the level anti-crossing in the spectrum of two dots at zero detuning. At the end of the sequence ($t = T$), the electron is ideally localized in the rightmost dot and remains in the ground orbital state. To simulate shuttling, we solved the 1D time-dependent Schr\"odinger equation numerically. Results for a 3-dot simulation are shown in figure~\ref{fig8}. For $T$ larger than a threshold value $T_{th}$, the simulated wavefunction has a large overlap with the ground state at all times. Non-adiabatic behaviour occurs for $T<T_{th}$ (see also section~\ref{sect: adiabaticity} in the Appendix). The state fidelity is defined as $|\langle \Psi_g(t) | \Psi(t)\rangle|^2$, where $\Psi_g(t)$ is the ground state wavefunction at time $t$, and $\Psi(t)$ is the actual state. The data in figure~\ref{fig8} indicate $T_{th}\approx 3$ ns. For shorter sequence times, the electronic wavefunction develops appreciable overlap with excited orbital states and is not properly localized in the target dot at the end of the sequence (see Appendix, figure~\ref{figA1}). For the non-adiabatic curves in figure~\ref{fig8}, the initial drop in fidelity occurs when dots 1 and 2 are near the resonant tunnelling point, where the energy gap between ground and excited states is determined by the tunnel coupling. A larger tunnel coupling allows for faster shuttling, although this is limited by the condition $\epsilon_t << E_g$, where $E_g$ is the energy gap between ground and first excited state in an isolated dot. Additional features can be seen in the middle of the sequence when $V_1$ and $V_3$ are swept rapidly, and near the second resonant tunnelling point. In our simulated dot, $E_g \approx 3$ meV, whereas valley splittings are typically a few hundred $\mu$eV in MOS dots. Thus, including the first excited valley state is expected to require slower sweeps to remain adiabatic; however, it is mainly the faster portions of the sweep that will be modified, as the slowest segments are still governed by $\epsilon_t$ (as long as $\epsilon_t << \Delta E_{vs}$). \\
\begin{figure}[t!h!]
\centering
\includegraphics[width = \linewidth]{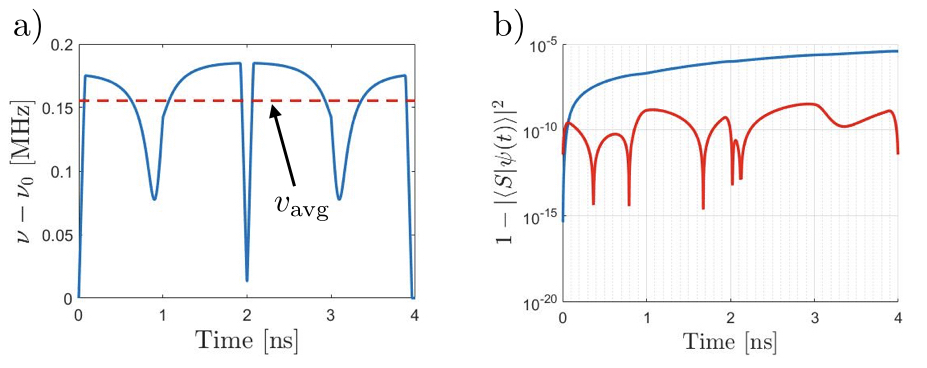}
\caption{\label{fig9} Stark effect and singlet phase rotation error for the 3-dot shuttling simulation. (a) Resonance frequency shift of the moving electron relative to its initial value $\nu_0 = 40$ GHz. The average value over the full duration, $\nu_{avg}$, is indicated by the red dotted line. (b) Phase error quantified as $1-F$, where $F = |\langle S | \psi(t)\rangle|^2$, $|S\rangle$ is the ideal singlet state and $\psi(t)$ is the actual spin state, a superposition of the singlet and $m_z = 0$ triplet states. The blue curve is for the case when the static electron has resonance frequency $\nu_0$. The red curve is the result of tuning the static dot so that its electron resonance frequency is $\nu_{avg}$, in which case the net phase error cancels out.}
\end{figure}
\subsection{Stark effect and singlet phase rotation error}
\indent A consequence of the weak (but non-zero) spin-orbit interaction in silicon together with a gate-induced local electric field is the Stark shift of the electronic $g$-factor \cite{veldhorst2014addressable, jock2018silicon}. The fractional variation of $g$ is typically of order $\sim 10^{-3}$ or less for practical gate voltages, but can be as large as $10^{-2}$. The normal component of the electric field, $E_z = -(\vec{\nabla} V)_z$, perturbs the $g$ value, which can be expressed as $\frac{\Delta g}{g} = \eta |E_z|^2$, where the Stark coefficient $\eta$ contains microscopic information and is normally determined experimentally \cite{rahman2009gate}. During the shuttling process, the electronic wavefunction experiences a time-dependent field $E_z$ which gives rise to modulation of $g$. For internode singlet distribution, this leads to errors since a difference in the $g$-factors of the static and moving electrons forming the singlet pair will cause a phase rotation of the state away from the singlet, towards the $m_z = 0$ triplet. In order to gauge the size of this error, we calculate the time-dependent $g$-factor of the moving dot using the instantaneous expectation value of the normal electric field, $\langle E_z(t) \rangle = \int dx \Psi^* E_z(x,t) \Psi$, with respect to the numerically calculated wavefunctions $\Psi(x,t)$. We take $\eta = 2.2$ (nm/V)$^2$ based on the empirical results reported in ref~\cite{veldhorst2014addressable}. Figure~\ref{fig9} shows the results from a 3-dot shuttling simulation with the same parameters used in figure~\ref{fig8}. Panel (a) shows the shift in resonance frequency $\nu = g \mu B_0/h$ as a function of time, where $\mu$ is the Bohr magneton and we take $B_0 = 1.43$ Tesla so that the initial resonance frequency is $\nu_0 = \nu(t=0) = 40$ GHz. The resonance frequency varies on a scale of $\sim 0.2$ MHz, with broad dips at the interdot tunneling transitions and abrupt changes corresponding to the large/fast voltage sweeps. Panel (b) shows the error accumulation $\epsilon = 1 - |\langle S |\psi \rangle|^2$ in terms of the overlap between the ideal singlet state $\kket{S}$ and the spin state at time $t$, $\kket{\psi(t)}$. The blue curve shows the case when the resonance frequency of the static qubit is $\nu_0$. The error surpasses $10^{-6}$ after these two shuttling steps. The net phase rotation error is coherent and due only to the offset between the \emph{average} value of $g$ for the moving dot versus the static value, and thus the error should increase with time as $\epsilon \sim 1 - \cos^2{(\pi(\Delta \nu) t)}$. For $\Delta \nu \approx 0.16$ MHz and $t = 4$ ns for two shuttle steps, this yields $\epsilon \approx 4\times10^{-6}$. For 30 shuttle steps (i.e. $1.74~\mu$m travel distance), this would correspond to an error of about 0.1$\%$. However, by tuning the $g$-factor of the static qubit to match $\nu_{avg}$ (red dotted line in panel (a)), the phase rotations are made to cancel over the course of the sequence and a much reduced error is obtained, as shown by the red curve in panel (b). In the latter case, the maximum error is $\sim 10^{-9}$, and returns to a negligibly small value at the end of the sequence. The Stark dephasing error can thus be mitigated with proper electrostatic tuning of the static qubit. Equivalently, one can think of the tuning correction as applying a small $\hat{z}$ rotation to one of the qubits to compensate for the net phase pickup of the shuttling sequence. \\
\indent From these simulations, we find that the modulation amplitude for $g$ scales roughly linearly with the range over which the gate voltages are swept, i.e. about 1 MHz/V. The voltage sweep range should thus be kept as low as possible to reduce the potential for Stark dephasing error. Fast noise in the electrostatic potential due to fluctuating charge defects or gate voltage noise from external sources should also be considered, as it would lead to irreversible dephasing of the singlet state. For an experimentally viable level of rms gate voltage noise of several $\mu$eV, the noise-induced fluctuation range for $g$ would be negligibly small, only $\sim 10$ Hz. Direct spin-orbit effects, on the other hand, are expected to produce larger errors. Spin-orbit coupling in a silicon 2DEG has been estimated to be of order $\sim 2 ~\mu$eV$\cdot$nm \cite{prada2011spin}, which yields a spin-orbit length $\sim 200~\mu$m. For an electron travel of 1.5$~\mu$m this would produce an error in the singlet fidelity $\sim 1.4 \times 10^{-4}$ due to spin rotation about a vector in the $\hat{x}-\hat{y}$ plane. Like the Stark effect phase rotation above, this is a coherent error, and is correctable by a suitable local rotation at the end of the process, in principle. Therefore, the \emph{average} error across a large ensemble of shuttled electrons (e.g. the many shuttling lines operating in parallel across the device) is correctable by local rotations, but the error spread due to non-uniformity of devices is not. The error spread, likely of the magnitude of the average error or less, should be tolerable by the surface code. We show in the next section that a threshold of nearly $1\%$ is obtained for dephasing error during shuttling when single and two-qubit gate errors are much smaller than the dephasing error. Multi-axis error such as weighted depolarizing noise during shuttling would likely have a lower threshold, however. Above-threshold errors would have to be mitigated by performing entanglement purification at the cost of additional ancilla qubits and gate operation overhead. We leave a more detailed analysis of the error mechanisms associated with spin shuttling to future work. \\
\subsection{Stabilizer repetition rate and other considerations}
\indent The time required to distribute a singlet between neighbouring nodes is $\sim \frac{L}{D}\tau$, where $L$ is the internode distance, $D$ is the dot dimension and $\tau$ is the time for a single shuttle operation. For a dot size of 50 nm and internode distance of 1.5 $\mu$m, the shuttling path consists of $\sim$ 30 dots. The shuttling parameters in figures ~\ref{fig8} and ~\ref{fig9} yield $\tau = 2$ ns, for a total time of 60 ns. Single-qubit ESR rotations typically require tens of nanoseconds at least, and with the equivalent of 16.5 $\pi$ rotations per subcycle, internode shuttling is not necessarily a bottleneck for the processor speed. With the inclusion of valley states and spin-orbit coupling, we expect the adiabaticity condition to be more stringent, reducing the attainable shuttling velocity. However, even if internode shuttling is an order of magnitude slower than our estimate above, the timescales of shuttling and intra-node operations would still be comparable. Reducing $\tau$ further should be possible with optimization of voltage sweep and dot parameters. Elongating the dots along the shuttling direction would reduce the number of dots required and could improve the operation fidelity. Since the orbital energy spacing decreases as $D^{-2}$, however, the adiabaticity condition will require slower operations as $D$ increases. Finding the optimal dot dimensions to maximize shuttling velocity while remaining adiabatic is worthy of investigation, along with designing optimal, smooth voltage sweep functions and optimizing tunnel rates. We emphasize that internode shuttling operations are global in that they proceed in parallel across the entire network for each stabilizer sequence (note, however, each subcycle of the surface code involves a distinct set of shuttle lines). Therefore, the electrodes controlling the shuttle path dots can be wired to common lines, assuming sufficiently high device uniformity. We expect that shuttling will be more tolerant of variations in dot parameters than qubit operations (i.e. as long as potential disorder is smaller than the minimum ground/excited state gap), although this should be investigated with numerical simulations. With common lines for shuttling, the number of external control wires can therefore remain manageable, of the order required for several plaquettes. We also emphasize that tuning the interdot tunnelling without explicit barrier gates relies only on the geometry of the dot gates and the voltage sequence, simplifying the device to a bare minimum of electrodes/wires.  \\
\indent The timescale for a full surface code cycle can be estimated by assuming realistic values for all operations in the subcycle of figure~\ref{fig4}. A full cycle consists of four subcycles (see Appendix~\ref{sect: scheduling}). The times we assume for singlet loading, internode shuttling, $\sqrt{SWAP}$ gates, emptying the ancilla dots and dispersive charge detection are 20 ns, 60 ns, 1 ns, 10 ns and 10 ns, respectively. Each subcycle has the equivalent of 16.5 $\pi$ rotations, including the control-$Z$ operation involved in each ancilla readout ($\hat{z}$ rotations are synthesized from $\hat{x},\hat{y}$ rotations). Single-qubit ESR gates therefore make the dominant contribution to the cycle time for Rabi frequencies below $\sim 100$ MHz. For an ESR Rabi frequency of (100, 10, 1) MHz, a full cycle requires approximately (1.2, 4.2, 33.9) $\mu$s. A plot of the full cycle rate versus ESR Rabi frequency is given in Appendix~\ref{sect: speed}. Although we have not considered logical qubit operations, which involve alternate stabilizer sequences on a subset of nodes, the four-qubit stabilizer rate should still give a reasonable estimate of processor speed for computation. The timescale for factoring a large number using Shor's algorithm has been estimated based on the surface code protocols for implementing logic gates described in ref.~\cite{fowler2012surface}. To factor a 2000-bit number in a scaled-up version of our network/node processor we estimate would require $\sim 23$ ($\sim 7$) days at 10 MHz (100 MHz) single-qubit Rabi frequency. \\
\indent Although we have so far assumed a global ESR field (e.g. placing the device chip inside a macroscopic microwave cavity), the highest Rabi frequencies are typically achieved with micromagnets \cite{yoneda2018quantum, obata2010coherent, watson2018programmable}. With the latter approach, direct control-$Z$ gates also become possible when the Zeeman energy difference between neighbouring qubits is comparable to the interdot tunnel coupling \cite{watson2018programmable}. These gates could be significantly faster than the standard control-$Z$ sequence we consider above, and potentially yield higher fidelities. Since micromagnets are not compatible with the singlet-triplet readout scheme proposed herein, a different method would be required, such as spin-dependent tunnelling to a reservoir together with fast charge sensing \cite{elzerman2004single}. This would eliminate the control-$Z$ gates used in the ancilla measurements in our scheme, potentially speeding up the processor. On the other hand, we expect that spin shuttling will be adversely affected by the presence of micromagnets, in general. As discussed in ref~\cite{zhaocoherent2018}, an inhomogeneous magnetic field along the interdot axis together with a valley splitting comparable to Zeeman energy can yield a high probability for spin flip during shuttling. The stray field along the spin quantization axis (external field direction) would also lead to significant phase rotation in the $m_z=0$ singlet-triplet subspace. However, since the micromagnet field is static and the shuttling voltage sequence can be fixed, the phase pickup at the end of the sequence is, in principle, correctable by an appropriate local $\hat{z}$ rotation. 
\section{Surface code error thresholds}
In quantum error correction, if the error rate of the physical components is below a certain threshold, the error rate of the logical qubits can be reduced by scaling up the code. The error threshold of surface codes is highly dependent on the way the stabilizer check circuit is implemented and the error models of the physical components. Its exact value can be obtained via simulations of the error correction circuit using the Gottesman-Knill theorem \cite{gottesman1998heisenberg, aaronson2004improved}. Assuming depolarizing noise for all the physical components, the error threshold of surface codes can take values between $0.5\% \sim 1\%$ under different circuit implementations \cite{stephens2014fault}. In our proposed quantum dot network architecture, $\sqrt{\text{SWAP}}$ is the basic building block of two-qubit gates instead of control-$Z$ or control-NOT gates. Failures of $\sqrt{\text{SWAP}}$ will predominantly lead to SWAP errors instead of depolarizing errors (see Appendix \ref{sect: swap_error}). For the shuttling process, we consider dephasing noise instead of depolarizing noise based on the findings of the previous section that phase rotation due to $g$-factor modulation should dominate the singlet state error. As mentioned above, although this is a coherent error for each shuttled electron, there is a spread in errors across the device, and this justifies the use of a dephasing model. The error types we consider in the stabilizer check circuit are the following:
\begin{itemize}
    \item Single-qubit gates, initialization and measurement: depolarizing errors with probability $p_{1q}$
    \begin{align*}
    \rho \rightarrow (1-p_{1q}) \rho + \frac{p_{1q}}{3}\left( X\rho X + Y\rho Y + Z\rho Z\right)
    \end{align*}
    \item $\sqrt{\text{SWAP}}$ gate: SWAP errors with probability $p_{swap}$
    \begin{align*}
    \rho \rightarrow (1-p_{swap}) \rho + p_{swap}\text{SWAP} \cdot\rho \cdot\text{SWAP}
    \end{align*}
    \item Shuttling process: dephasing errors (due to $g$-factor modulation) with probability $p_{sh}$
    \begin{align*}
    \rho \rightarrow (1-p_{sh}) \rho + p_{sh}Z\rho Z
    \end{align*}
\end{itemize}
In particular, SWAP errors and the fact that $\sqrt{\text{SWAP}}$ is non-Clifford will give rise to non-Pauli noise in the circuit (see Appendix \ref{sect:CZ_error}). To efficiently simulate the error correction circuit using the Gottesman-Knill theorem, non-Pauli error operators must be converted into Pauli noise by twirling.
\subsection{Twirling}
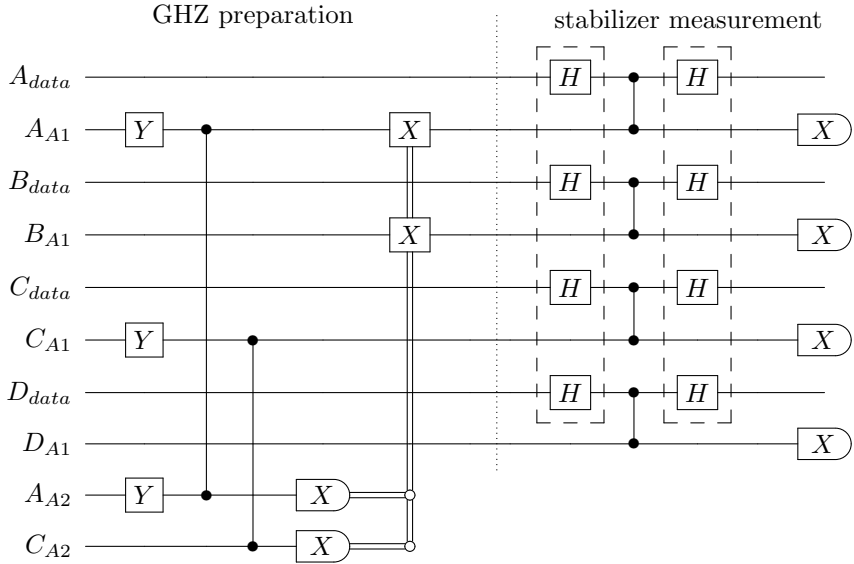
\begin{figure}[htbp]
    \centering
    \begin{align*}
    \Qcircuit @C=1.5em @R=.7em {
        &&&& \ustick{\text{GHZ preparation}} &&&&\ar@{.}[]+<-0.5em,1em>;[dddddddd]+<-0.5em,-1em>&&&\ustick{\text{ stabilizer measurement\quad}}&&&\\
        &\lstick{A_{data}}&\qw     &\qw     &\qw&\qw&\qw&\qw&\qw&\gate{H}&\ctrl{1}&\gate{H}&\qw&\qw\\
        &\lstick{A_{A1}} &\gate{Y}&\ctrl{7}&\qw     &\qw         &\gate{X}&\qw&\qw&\qw&\ctrl{0}&\qw&\qw&\measureD{X}\\
        &\lstick{B_{data}}&\qw     &\qw     &\qw     &\qw         &\qw&\qw&\qw&\gate{H}&\ctrl{1}&\gate{H}&\qw&\qw\\
        &\lstick{B_{A1}} &\qw     &\qw     &\qw     &\qw         &\gate{X}\cwx[-2]&\qw&\qw&\qw&\ctrl{0}&\qw&\qw&\measureD{X}\\
        &\lstick{C_{data}}&\qw     &\qw     &\qw     &\qw         &\qw&\qw&\qw&\gate{H}&\ctrl{1}&\gate{H}&\qw&\qw\\
        &\lstick{C_{A1}} &\gate{Y}&\qw     &\ctrl{4}&\qw         &\qw&\qw&\qw&\qw&\ctrl{0}&\qw&\qw&\measureD{X}\\
        &\lstick{D_{data}}&\qw     &\qw     &\qw     &\qw         &\qw&\qw&\qw&\gate{H}&\ctrl{1}&\gate{H}&\qw&\qw\\
        &\lstick{D_{A1}} &\qw     &\qw     &\qw     &\qw         &\qw&\qw&\qw&\qw&\ctrl{0}&\qw&\qw&\measureD{X}\\
        &\lstick{A_{A2}} &\gate{Y}&\ctrl{0}&\qw     &\measureD{X}&\cctrlo{-5}\\
        &\lstick{C_{A2}} &\qw     &\qw     &\ctrl{0}&\measureD{X}&\cctrlo{-1}
        \gategroup{2}{10}{8}{10}{1em}{--} \gategroup{2}{12}{8}{12}{1em}{--}
    }
    \end{align*}
    \caption{Simplified diagram of the stabilizer check circuit. Control-$Z$ gates are indicated by the vertical lines connecting dots. The following input pairs are initialized as singlets: ($A_{A1}$, $B_{A1}$), ($C_{A1}$, $D_{A1}$), ($A_{A2}$, $C_{A2}$). The Hadamard ($H$) gates in the dashed boxes are used only for the $X$ stabilizer. When the parity of the measurement results on $A_{A2}$ and $C_{A2}$ is odd, two additional $X$ gates are applied on $A_{A1}$, $B_{A1}$ to produce the GHZ state.}
    \label{Fig:stb_circuit}
\end{figure}

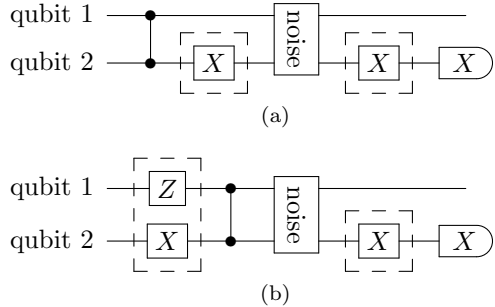
\begin{figure}[htbp]
    \centering
    \subfloat[][]{
        \Qcircuit @C=1.5em @R=.7em {
            &\lstick{\text{qubit 1}} &\ctrl{1} &\qw    &\multigate{1}{\rotatebox{270}{noise}}    &\qw        &\qw \\
            &\lstick{\text{qubit 2}} &\ctrl{0} &\gate{X} &\ghost{\rotatebox{270}{noise}}     &\gate{X}  &\measureD{X}
            \gategroup{2}{4}{2}{4}{1em}{--} \gategroup{2}{6}{2}{6}{1em}{--}
        }
    }
    \qquad
    \qquad
    \qquad
    \subfloat[][]{
        \Qcircuit @C=1.5em @R=.7em {
            &\lstick{\text{qubit 1}}&\gate{Z} &\ctrl{1}     &\multigate{1}{\rotatebox{270}{noise}}    &\qw      &\qw    \\
            &\lstick{\text{qubit 2}}&\gate{X} &\ctrl{0} &\ghost{\rotatebox{270}{noise}}     &\gate{X} &\measureD{X}
            \gategroup{1}{3}{2}{3}{1em}{--} \gategroup{2}{6}{2}{6}{1em}{--}
        }
    }
    \caption{(a) If it were the case that the noise existed as a separate physical process from the ideal gate, then one could target the noise directly by twirling operations (shown in dashed boxes). (b) However, in reality the noise process is inseparable from the ideal gate, therefore we permute one of the twirling operations back through that gate to obtain the physically achievable twirling protocol. }
    \label{Fig:CZ_twirl_full}
\end{figure}

Twirling is used for converting arbitrary error channels into Pauli channels by conjugating the noise with Pauli gates randomly  chosen from the twirling gate set~\cite{cai2018constructing}. The Pauli channel we obtain is the incoherent superposition of the Pauli basis of the original noise. For example, after twirling, a swap noise $\text{SWAP} = \frac{1}{2}\left(I_1I_2+X_1X_2+Y_1Y_2+Z_1Z_2\right)$ will be transformed into $ \frac{1}{4}\left(\widehat{I_1I_2}+\widehat{X_1X_2}+\widehat{Y_1Y_2}+\widehat{Z_1Z_2}\right)$, where $\widehat{}$ denotes a super-operator\footnote{For example, $\left(\widehat{A} + \widehat{B}\right) \rho = A\rho A^\dagger + B\rho B^\dagger$}. Twirling is proven to be effective in error threshold simulations~\cite{geller2013efficient, gutierrez2015comparison}.\\
\indent To run error threshold simulations, we must first obtain the error distribution for each round of the stabilizer check. This can be obtained via a full quantum simulation of the stabilizer check circuit, which is shown in figure~\ref{Fig:stb_circuit}. In this circuit, the non-Pauli errors arise from the failures of the elements comprising the control-$Z$ gates (Appendix \ref{sect:CZ_error}). Using conventional twirling on these two-qubit errors requires the full Pauli set of the size $4^2 = 16$ as the twirling gate set~\cite{bennett1996mixed, dur2005standard}. Hence, if we want the exact error distribution for each round of stabilizer check, $16^6$ possibilities must be iterated over since there are $6$ control-$Z$ gates in the circuit. 
In the stabilizer check circuit, control-$Z$ gates are always followed by an $X$ measurement whose results are forgotten at the end; we only record the parity of the $X$ measurements in both the preparation and the stabilizer check stage. Using the method proposed in \cite{cai2018constructing}, it is found that the gate set $\{I, X_2\}$ of size $2$ is sufficient for twirling. Thus, we only need to iterate over $2^6$ possibilities instead of $16^6 = 2^{24}$ to obtain the exact error distribution. The twirling circuit is shown in figure~\ref{Fig:CZ_twirl_full}.

\begin{figure}[h!]
    \centering
    \includegraphics[scale = 0.7]{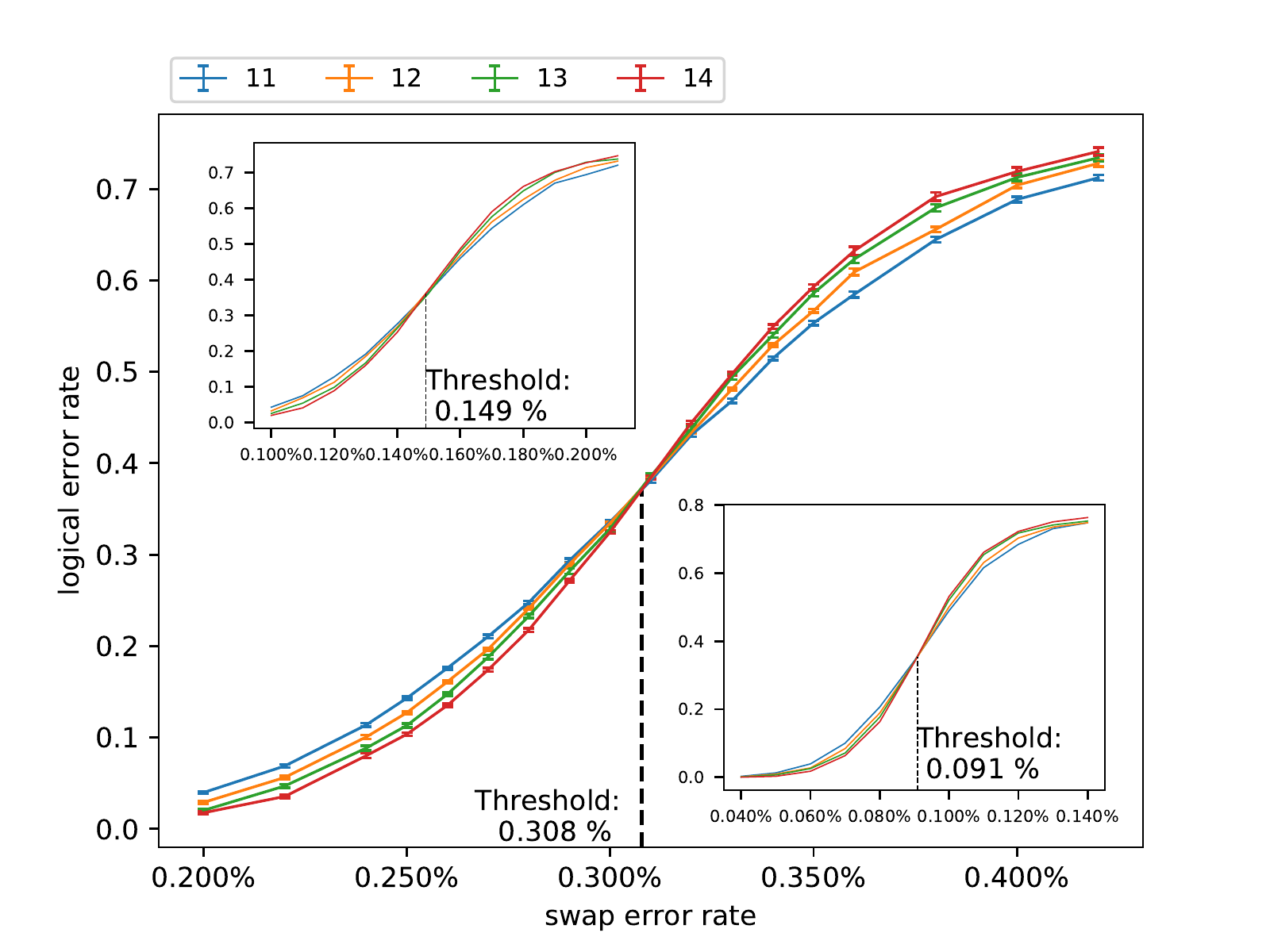}
    \caption{Fault tolerance thresholds with respect to error in $\sqrt{\text{SWAP}}$ gates, with $p_{sh}=0$ and different ratios $\frac{p_{1q}}{p_{swap}}$. 
        \textbf{Main plot:} $\frac{p_{1q}}{p_{swap}} = 0.1$.
        \textbf{Top left:} $\frac{p_{1q}}{p_{swap}} = 0.5$.
        \textbf{Bottom right:} $\frac{p_{1q}}{p_{swap}} = 1.0$. 
        The legend shows curves of different colours corresponding to different code distances $d$. The corresponding number of nodes (data qubits) in our network is $d^2+(d-1)^2$. The dashed lines indicate the threshold values.}
    \label{Fig:02}
\end{figure}
\subsection{Threshold simulation results}
\begin{figure}[h!]
    \centering
    \includegraphics[scale = 0.7]{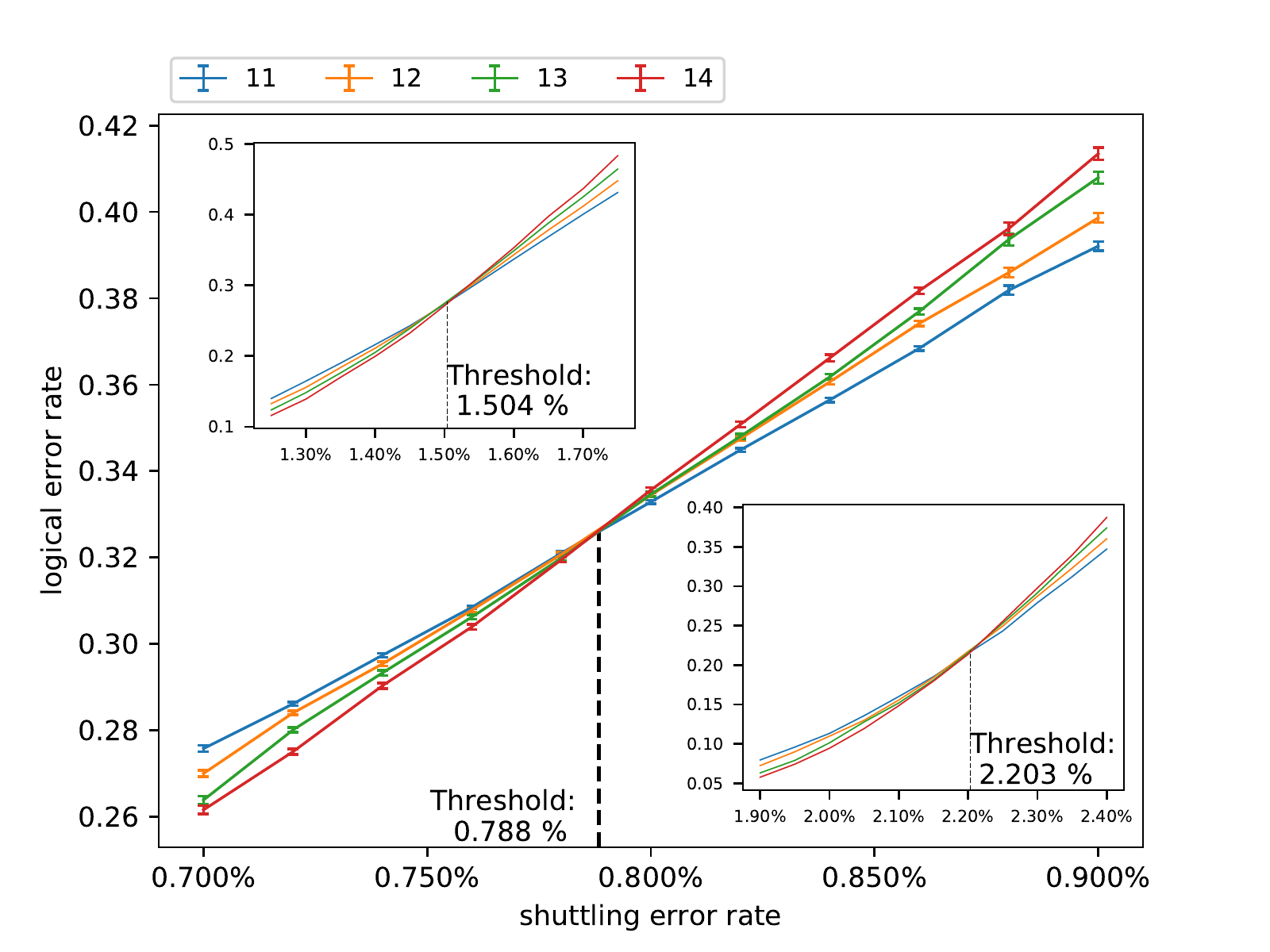}
    \caption{Fault tolerance thresholds with respect to dephasing errors during shuttling, with $\frac{p_{1q}}{p_{swap}}=0.1$ and different values of $p_{swap}$. 
        \textbf{Main plot:} $p_{swap} = 0.2\%$.
        \textbf{Top left:}$p_{swap} = 0.1\%$.
        \textbf{Bottom right:}  $p_{swap} = 0$.
          The legend shows curves of different colours corresponding to different code distances $d$. The dashed lines indicate the threshold values.}
    \label{Fig:03}
\end{figure}

If we assume no dephasing error during shuttling ($p_{sh} = 0$), and fix the error ratio between single-qubit errors ($p_{1q}$) and SWAP errors ($p_{swap}$), we obtain the threshold plots for $p_{swap}$ as shown in figure~\ref{Fig:02}. The threshold is defined as the gate error rate at which there is a crossover between the logical error rate increasing with code size (above threshold) and the error rate decreasing with code size (below threshold). The threshold can be seen here for code distances ranging from $d=11$ to 14, where the corresponding number of nodes (or data qubits) in our network is $d^2+(d-1)^2$. Single-qubit operations are typically achieved with higher fidelity than two-qubit operations. Under the realistic assumption that $\frac{p_{1q}}{p_{swap}} = 0.1$, we obtain a threshold of $0.31\%$ for $p_{swap}$, which is of the same order as the threshold for a depolarizing noise model ($0.5\% \sim 1\%$~\cite{stephens2014fault}). Keeping the assumption that $\frac{p_{1q}}{p_{swap}} = 0.1$, and fixing $p_{swap}$ to a below-threshold value $0.2\%$, we obtain the threshold of the shuttling dephasing error to be $0.79\%$ as shown in  figure~\ref{Fig:03}. This threshold approaches $2\%$ in the limit $p_{swap}, p_{1q}\rightarrow 0$. This shows a relatively high tolerance of the surface code to spin dephasing errors during internode shuttling. As noted in section~\ref{sect:stabilizer}, each $Z$ and $X$ stabilizer is split into two sequential operations since any two neighbouring 4-node plaquettes cannot be stabilized simultaneously. In these simulations, we have neglected this by assuming that idle data qubits decohere at a much slower rate than those experiencing active gate operations. \\
\indent Finally, we note that demonstrated electron spin coherence times in MOS dots (with isotopic enhancement to remove $^{29}$Si nuclear spins) are compatible with fault tolerance in our architecture. Ref.~\cite{veldhorst2015two} reports a dephasing time $T^*_2 \sim 120 \mu$s and $T_2 \sim 28$ ms under CPMG refocusing. This should be compared to our estimated stabilizer cycle time $\sim 2 \mu$s. Thus, with refocusing, the probability of a phase flip error purely due to $T_2$ is of order $10^{-4}$ per cycle, which should be well below the fault tolerance threshold for the surface code. \\
\section{Conclusions}
In summary, we have proposed a surface code realization for quantum dot spin qubits in silicon based on a network of nodes. The spatial separation of the nodes allows data qubits to be better isolated and will ease constraints on wiring density and integration of classical circuit elements to support control and readout functions. As each node contains fewer than 10 quantum dots, demonstrating a fully functional node is nearly within the grasp of current technology. Connecting nodes relies on shuttling of electrons over medium-range distances ($\sim 1~\mu$m) and maintaining the fidelity of the distributed spin singlet states. We find value in separating the scaling problem into these two streams - local operations and entanglement distribution -  that can be developed in parallel. Realistic simulations with the simplest possible gate electrode geometry show that adiabatic shuttling can be realized on timescales that do not necessarily present a speed bottleneck to the processor. Simulations suggest that the dominant error in a clean system is uncontrolled phase rotation due to the modulation of the electronic $g$-factor during shuttling, owing to the Stark effect. While this error $\sim 0.1\%$ may be tolerable by a scaled-up surface code, we show how it can be much further reduced by appropriate tuning of the stationary electron's $g$-factor. These shuttling results, however, do not tell the whole story because we have not included multiple valleys, direct spin-orbit coupling or charge state decoherence in the simulations. The combined effects of these factors could indeed make coherent spin transport over many dots difficult or impossible, and it is critical that simulations based on realistic gate geometries like ours be extended to take these into account. While we have chosen to focus on spin shuttling in this paper, of course, any viable method for internode entanglement distribution can be used in its place. \\
\indent Achieving fault tolerance is a critical goal for a scalable processor. Using reasonable noise models, we estimate error thresholds with respect to single and two-qubit gate fidelities as well as dephasing errors due to shuttling. A twirling protocol allows us to transform the non-Pauli noise associated with exchange gate operations into Pauli noise, making it possible to use the Gottesman-Knill theorem to efficiently simulate large codes. Not surprisingly, the surface code is found to be more robust to singlet dephasing errors than to errors in $\sqrt{\text{SWAP}}$ operations. A $\sqrt{\text{SWAP}}$ error threshold of $0.31\%$ was found when the probability of single-qubit error is 0.1 times that of the two-qubit exchange gate. A dephasing (shuttling) threshold of 0.79$\%$ was found when $\sqrt{\text{SWAP}}$ and single-qubit error probabilities are 0.2$\%$ and 0.02$\%$, respectively. Thus, compared to the current state of the art in silicon spin qubits \cite{watson2018programmable, veldhorst2015two, eng2015isotopically, yoneda2018quantum}, both single-qubit and two-qubit gate infidelities must be reduced by at least an order of magnitude to achieve fault tolerant levels (of course, this statement applies equally well to other realizations and error correction schemes). The error models used to estimate fault tolerance thresholds will become more realistic as they are further informed by experiments at progressively finer levels of control. We also expect that the uniformity of tuning parameters/properties of nominally identical dots must improve by at least an order of magnitude compared to what has been demonstrated experimentally so far \cite{zajac2016scalable}. This is so that shared control lines, a practical necessity for scalability, can be feasible. Taking advantage of the internode spacing in our architecture, we envision that local floating gate electrodes could be programmed to apply small electrostatic corrections to the quantum dots forming the nodes, allowing control pulse sequences to be applied globally. On the other hand, we expect that electron shuttling can be made robust to sufficiently small variations in dot uniformity, so that shared global control of spin transport will be feasible without the need for correction gates. The robustness of shuttling operations is a subject for future work, including a more detailed study of spin-orbit and valley effects.\\
\indent Similar to ref.~\cite{nickerson2013topological}, we have only considered the case that all nodes perform four-qubit stabilizer operations, which is equivalent to logical qubit storage rather than computation. It is expected that error thresholds for computation will be similar, since the four-qubit stabilizer constitutes the bulk of operations and alternative stabilizers needed for computation are only required at boundaries. It remains to determine the precise operations within boundary nodes during computations. Clearly, we must have the ability to address boundary nodes individually, as well as the bulk nodes collectively, noting that the boundaries move during computations and thus can involve many, if not all, nodes at some point in the computation. Individual addressing of nodes will also be required during initial calibration, e.g. for setting the values of correction floating gates. An appropriate multiplexing scheme utilizing conventional transistor circuits as in ref.~\cite{veldhorst2017silicon} could be applied, noting that our scheme can make available enough space for 3D interconnects using present-day CMOS technologies (power dissipation at mK temperatures remains a challenge). Performing massively parallel readout operations in any surface code architecture is another challenge for which relatively little has been discussed in literature. Both time and frequency multiplexing can be used with RF reflectometry, but it is not yet obvious how this can be done at large scale while keeping measurement latency within acceptable bounds \cite{chamberland2018fault}.
\section*{Acknowledgements}
This work was supported by Natural Sciences and Engineering Research Council of Canada (NSERC). ZC acknowledges support from Quantum Motion Technologies Ltd. SCB acknowledges support from EPSRC grant EP/M013243/1 (NQIT Quantum Hub). We thank Francois Sfigakis, Emma Bergeron and Kaveh Gharavi for helpful discussions.  \\
\section{Appendix}
\subsection{4-qubit GHZ state}
\label{sect:GHZ}
Here we outline the mathematical details underlying the circuit of figure~\ref{fig4}. In our notation, commas separate nodes from each other as $\ket{A,B,C,D}$.  Within a particular node, the first qubit represents the A1 qubit and the second (if written) indicates the A2 qubit. A blank space in the A2 location means that qubit is not present. 

The initial state after the singlet distributions between A1 in both node pairs A-B and C-D (segment 1 in figure~\ref{fig2}) is
\begin{align}
\ket{\Psi} &= \frac{1}{\sqrt{2}}\Big(\ket{0,1}_{AB} - \ket{1,0}_{AB}\Big)\otimes \frac{1}{\sqrt{2}}\Big(\ket{0,1}_{CD} - \ket{1,0}_{CD}\Big) \\
&= \frac{1}{2}\Big(\ket{0,1,0,1} - \ket{0,1,1,0} - \ket{1,0,0,1} + \ket{1,0,1,0}\Big)
\end{align}
Next, the singlet distribution between A2 in the nodes A-C (segment 2 in figure~\ref{fig2}) gives the state
\begin{align}
\ket{\Psi} &= \frac{1}{2\sqrt{2}}\Big(\ket{00,1,01,1} - \ket{00,1,11,0} - \ket{10,0,01,1} + \ket{10,0,11,0} \\
&-\ket{01,1,00,1} + \ket{01,1,10,0} + \ket{11,0,00,1} - \ket{11,0,10,0}\Big)
\end{align}
The first step of segment 3 in figure~\ref{fig2} is a $Y_{\pi}$ rotation on the A1 qubits in nodes A and C and on the A2 qubit in A.  This transforms the singlets into the $\ket{\Phi^+}$ Bell states, and the total state becomes
\begin{align}
\ket{\Psi} &= \frac{-i}{2\sqrt{2}}\Big(\ket{11,1,11,1} + \ket{11,1,01,0} + \ket{01,0,11,1} + \ket{01,0,01,0} \\
&+\ket{10,1,10,1} + \ket{10,1,00,0} + \ket{00,0,10,1} + \ket{00,0,00,0}\Big)
\end{align}
Next, we perform a control-$Z$ operation between the A1 and A2 qubits in nodes A and C,
\begin{align}
\ket{\Psi} &= \frac{-i}{2\sqrt{2}}\Big(\ket{11,1,11,1} - \ket{11,1,01,0} - \ket{01,0,11,1} + \ket{01,0,01,0} \\
&+\ket{10,1,10,1} + \ket{10,1,00,0} + \ket{00,0,10,1} + \ket{00,0,00,0}\Big)
\end{align}
To create the GHZ state distributed across all four A1 qubits, we perform an $X$ basis measurement on the A2 qubits in nodes A and C.  Rewriting the state above in the $X$ basis, it is easy to find the state of the A1 qubits conditional on the four measurement outcomes: 
\begin{align}
+_A,+_C &\rightarrow \frac{1}{\sqrt{2}}\Big(\ket{0,0,0,0} + \ket{1,1,1,1}\Big) \,\,\,\,\,\,\, -_A,-_C \rightarrow \frac{1}{\sqrt{2}}\Big(\ket{0,0,0,0} + \ket{1,1,1,1}\Big) \\
+_A,-_C &\rightarrow \frac{1}{\sqrt{2}}\Big(\ket{0,0,1,1} + \ket{1,1,0,0}\Big) \,\,\,\,\,\,\, -_A,+_C \rightarrow \frac{1}{\sqrt{2}}\Big(\ket{0,0,1,1} + \ket{1,1,0,0}\Big)
\end{align}
Since the four outcomes above occur with equal probability, the even parity outcomes give the GHZ state with probability 1/2, while the odd parity outcomes give a state with equal probability that is transformed into the GHZ state by applying an $X_\pi$ rotation on the A1 qubits in two of the nodes. The four-qubit GHZ state preparation is therefore deterministic. \\

\subsection{Sequence of four-qubit stabilizer operations}
\begin{figure}[h!]
\centering
\includegraphics[width = 0.8\linewidth]{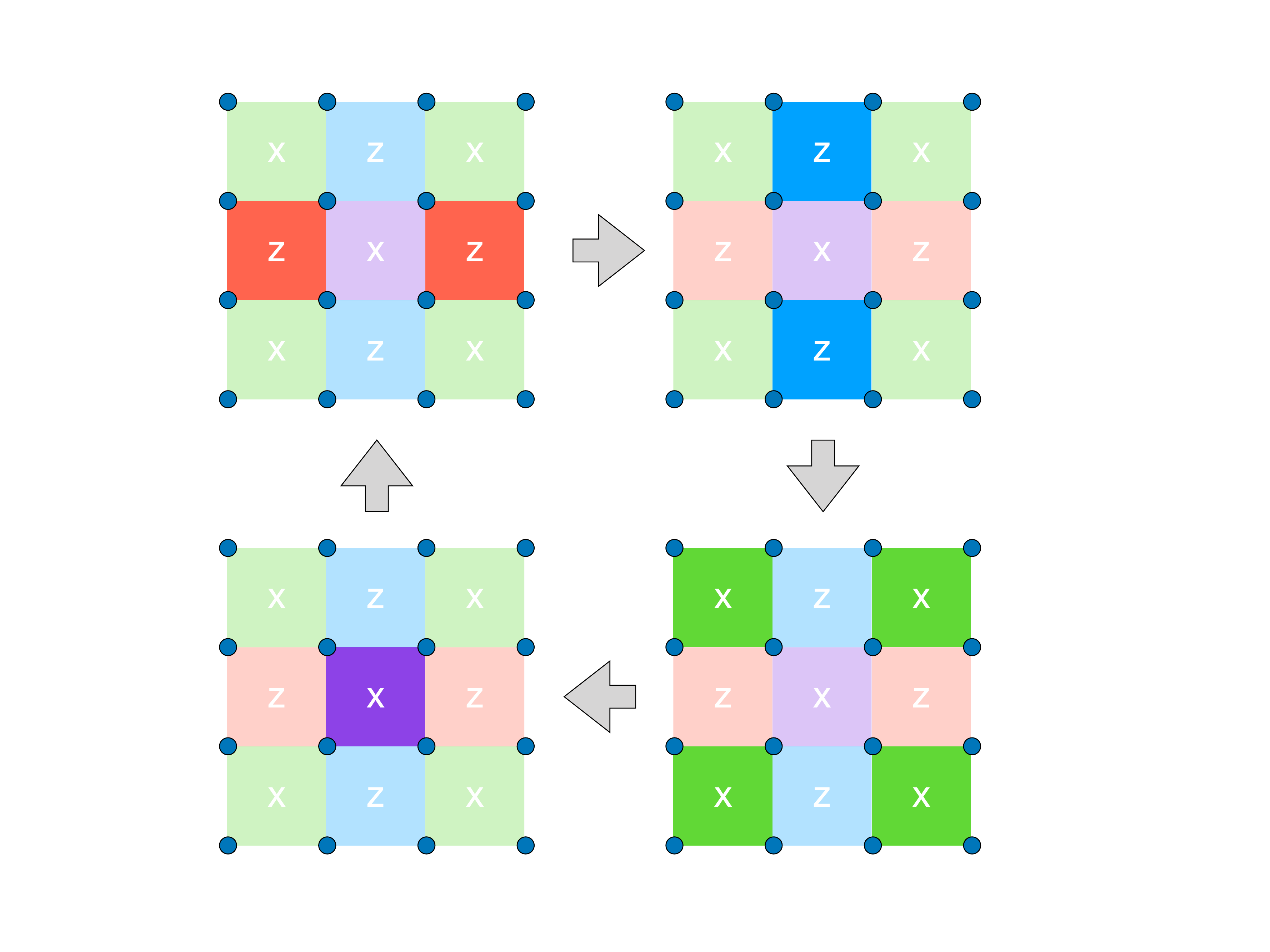}
\caption{\label{schedulefig} A full stabilizer cycle consists of the four steps indicated in the figure, since adjacent 4-node plaquettes cannot be stabilized at the same time. Starting at the upper left, the $Z$ stabilizer is split into two steps, followed by the two-step $X$ stabilizer. The cycle then repeats. }
\end{figure}

See figure~\ref{schedulefig}. 
\label{sect: scheduling}

\subsection{Adiabaticity of the electronic wavefunction during shuttling}\label{sect: adiabaticity}
\subsubsection{Voltage sequence design and fine tuning}\label{sect: finetune}
The gate voltage sequence for shuttling, e.g. the example in figure~\ref{fig8}, was designed using a set of 1D potentials calculated using the Nextnano software \cite{birner2007nextnano}. Approximately 1000 potentials were calculated based on gate voltage increments of 0.01 or 0.02 V. Potential landscapes at a finer gate voltage resolution were obtained by linear interpolation. Coherent evolution under the time-dependent Schr\"odinger equation was calculated using a time step of $5\times 10^{-17}$ s, e.g. $\sim 10^8$ time points for a 4 ns sequence. Consider a two-dot system with local gate voltages $V_1$ and $V_2$, the electron initialized in dot 1 and $V_1 > V_2$. To transfer the electron to dot 2, we first sweep $V_2$ in the positive direction while holding $V_1$ fixed. This can be done very quickly over the range of $V_2$ for which the wavefunction localized in dot 1 is insensitive to $V_2$. This fast sweep ends when interdot tunnelling `turns on' and there is a small probability for the electron to be in dot 2; we chose an arbitrary threshold of $\sim 0.1\%$ probability. $V_2$ is then swept slowly enough to continue to satisfy the approximate adiabaticity condition \cite{comparat2009general}
\begin{align}
\mathlarger{\sum}_{m\neq g} \frac{\langle\psi_m\kket{\dot{\psi}_g}}{|E_m - E_g|} << 1, 
\end{align}
where $\dot{\psi}_g$ is the time derivative of the instantaneous ground state $\psi_g(t)$ and $\psi_m$ is the $m^{\text{th}}$ excited state orbital. As the resonant tunnelling point $V_2 \simeq V_1$ is approached, the relevant energy gap for adiabaticity is given by $2\epsilon_t$ where $\epsilon_t$ is the tunnelling energy. $V_1$ is then swept slowly in the negative direction with $V_2$ fixed, until the tunnel coupling is sufficiently `off' that $V_1$ can be swept quickly without affecting the wavefunction now localized in dot 2. In a linear dot array, cross-capacitances between gates affect the dot potentials so that the exact resonant tunnelling points differ slightly from the points $V_j=V_k$ for adjacent dots $j$ and $k$. The correct resonant tunnelling points are identified by the electron having equal probability to be in both dots, and this is taken into account in the construction of the gate voltage sequences. 

\subsubsection{3-dot simulation results}
\begin{figure}[h!]
\centering
\includegraphics[width = 0.9\linewidth]{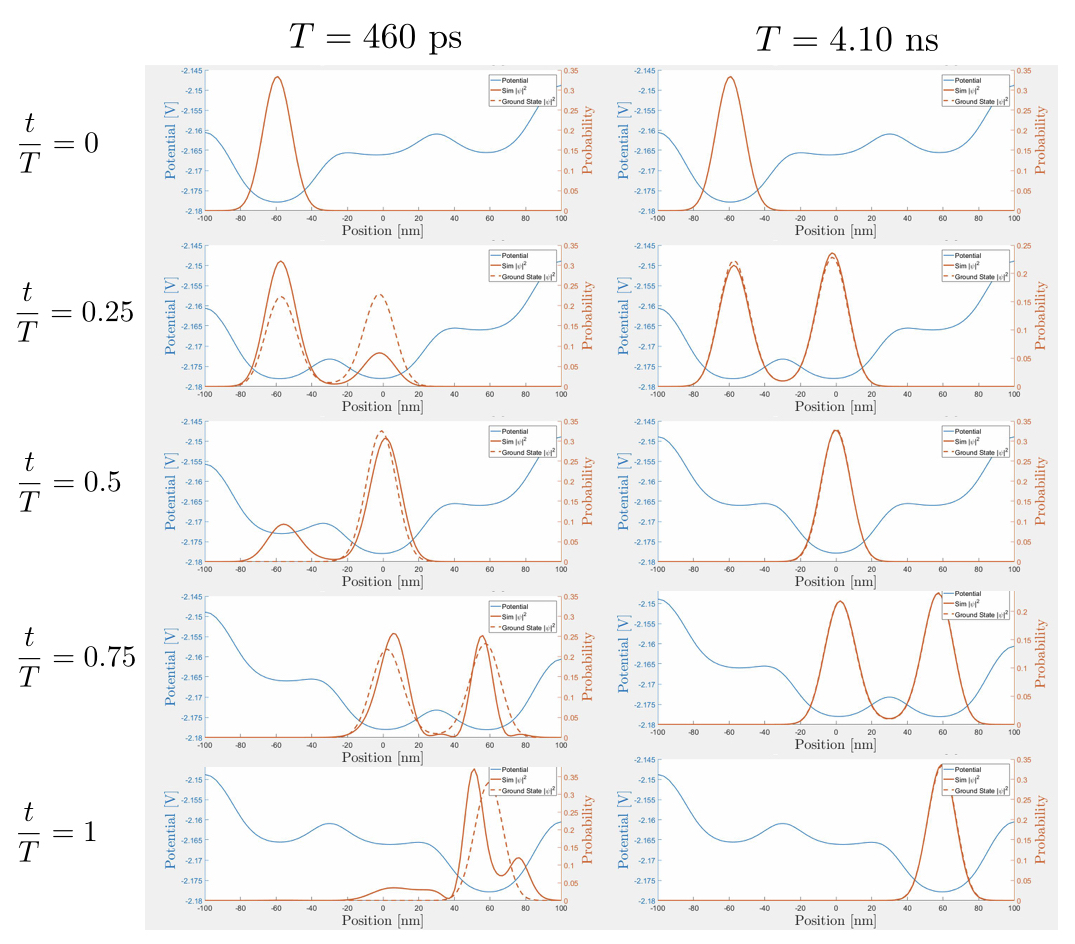}
\caption{\label{figA1} Comparison of 3-dot electron shuttling simulations showing adiabatic (right) versus non-adiabatic (left) behaviour. The left (right) panels correspond to a total sequence time $T = 0.46$ ns ($T = 4.1$ ns), respectively. Snapshots of the electron probability density are shown at different time points in the sequence (labels in the vertical column at left). The ideal ground state wavefunction is indicated by the dashed line and the actual wavefunction by the solid line.}
\end{figure}
Figure~\ref{figA1} shows results of the same 3-dot simulation reported in figure~\ref{fig8}. It shows the wavefunction evolution in two cases, where the total shuttle time is above (4.1 ns) and below (0.46 ns) the adiabaticity threshold of approximately 3 ns. For the shorter time, it is clear that by the end of the sequence the wavefunction (solid line) deviates significantly from the ideal ground state (dotted line), and has less than unit probability to be found in the target (third) dot. We expect that the adiabaticity threshold time can be reduced by increasing the tunnel coupling and by a more optimal voltage sequence design using smooth waveforms. 

\subsection{Processor speed}
\label{sect: speed}
\begin{figure}[h!]
\centering
\includegraphics[width = 0.75\linewidth]{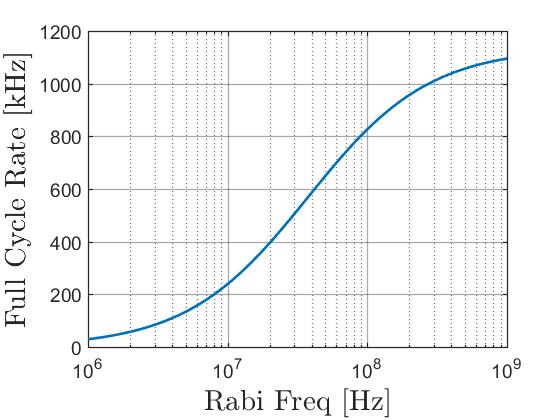}
\caption{\label{speed} Rate for the full stabilizer cycle (both $X$ and $Z$ stabilizers) versus the single-qubit control Rabi frequency. Below $\sim 100$ MHz, the cycle rate is dominated by single-qubit operations and is a linear function of the Rabi frequency. Above $\sim 100$ MHz, the cycle rate reaches a plateau limited by the durations of all other operations. Here we assume the following operation times: singlet loading $= 20$ ns, internode shuttling $= 60$ ns, $\sqrt{SWAP}$ gate $= 1$ ns, emptying ancilla dots $= 10$ ns and gate-dispersive charge detection $= 10$ ns. }
\end{figure}
See figure~\ref{speed}. 

\subsection{Fault tolerance threshold calculations: non-Pauli errors}
\subsubsection{Error of $\sqrt{\text{SWAP}}$} \label{sect: swap_error}
The evolution operator of the exchange interaction is:
\begin{align*}
\hat{U}_{ex}(\tau)  =  \begin{pmatrix}
e^{-iJ\tau/2}&0&0&0\\0&\cos(\frac{J\tau}{2})&-i\sin(\frac{J\tau}{2})&0\\0&-i\sin(\frac{J\tau}{2})&\cos(\frac{J\tau}{2})&0\\0&0&0&e^{-iJ\tau/2}
\end{pmatrix}\begin{matrix}
\ket{\uparrow\uparrow}\\\ket{\uparrow\downarrow}\\\ket{\downarrow\uparrow}\\\ket{\downarrow\downarrow}
\end{matrix}
\end{align*}
where $J$ is the exchange energy. If re-write $\theta = \frac{J\tau}{2}$, we have
\begin{align*}
\hat{U}_{ex}(\theta) = \cos(\theta) I - i\sin(\theta) \text{SWAP}
\end{align*}
A SWAP gate corresponds to $\theta = \pi/2$, and a $\sqrt{\text{SWAP}}$ gate to $\theta = \pi/4$. However, over- and under-rotations of exchange occur in experiments due to imprecise pulse timing or fluctuation of exchange strength due to charge fluctuations. If there is a $50\%$ percent chance of over and under rotation by $\epsilon\ll 1$ when we want to achieve a $\sqrt{\text{SWAP}}$ gate, we will have:
\begin{align*}
\hat{U}_{ex}(\frac{\pi}{4} \pm \epsilon) &= \cos(\frac{\pi}{4} \pm \epsilon) I - i\sin(\frac{\pi}{4} \pm \epsilon) \text{SWAP}\\
& = \frac{1}{\sqrt{2}}( I \pm i \text{SWAP}) - \frac{\epsilon}{\sqrt{2}}( I \pm i \text{SWAP}) \\
& = \pm\hat{U}_{ex}(\frac{\pi}{4}) \mp \epsilon \hat{U}_{ex}(\frac{3\pi}{4})\\
\end{align*}
Then the effective operation is
\begin{align*}
&\quad\frac{1}{2}\hat{U}_{ex}(\frac{\pi}{4} + \epsilon) \rho \hat{U}_{ex}^\dagger(\frac{\pi}{4} + \epsilon) + \frac{1}{2} \hat{U}_{ex}(\frac{\pi}{4} - \epsilon) \rho \hat{U}_{ex}^\dagger(\frac{\pi}{4} - \epsilon)\\
&=\hat{U}_{ex}(\frac{\pi}{4}) \rho \hat{U}_{ex}^\dagger(\frac{\pi}{4}) + \epsilon^2 \hat{U}_{ex}(\frac{3\pi}{4}) \rho \hat{U}_{ex}^\dagger(\frac{3\pi}{4})\\
& = \sqrt{\text{SWAP}} \rho \sqrt{\text{SWAP}}^\dagger + \epsilon^2 \left(\text{SWAP}\sqrt{\text{SWAP}}\right) \rho \left(\text{SWAP}\sqrt{\text{SWAP}} \right)^\dagger
\end{align*}
i.e. we have either perfect $\sqrt{\text{SWAP}}$ or $\epsilon^2$ probability of having a $\text{SWAP}$ error on top of $\sqrt{\text{SWAP}}$. Similar arguments can be applied to other symmetric over/under-rotation distributions that center on the correct rotation angles.

\subsubsection{Errors associated with the control-$Z$ gate}\label{sect:CZ_error}
The control-$Z$ gates are implemented as
\begin{align*}
\Qcircuit @C=1.5em @R=1.3em {
    &\lstick{\text{qubit 1}}&\gate{Z_{\frac{\pi}{2}}} &\multigate{1}{\rotatebox{270}{$\sqrt{\text{SWAP}}$}}    &\gate{Z_\pi} &\multigate{1}{\rotatebox{270}{$\sqrt{\text{SWAP}}$}}         &\qw\\
    &\lstick{\text{qubit 2}}&\gate{Z_{-\frac{\pi}{2}}} &\ghost{\rotatebox{270}{$\sqrt{\text{SWAP}}$}}     &\qw&\ghost{\rotatebox{270}{$\sqrt{\text{SWAP}}$}} &\qw\\
}
\end{align*}

There are three types of non-Pauli noise that can occur:
\begin{itemize}
    \item SWAP error after second $\sqrt{\text{SWAP}}$ $\Rightarrow$ SWAP error after the control-$Z$
    
    \item SWAP error after first $\sqrt{\text{SWAP}}$ $\Rightarrow$ ($Z_1Z_2$)$\cdot$SWAP error after the control-$Z$
    
    \item $\sigma^i_1$ error after $Z_{\pi}$ $\Rightarrow$ $\sqrt{\text{SWAP}} \cdot \sigma^i_1 \cdot \sqrt{\text{SWAP}}^\dagger$ error after the control-$Z$.
\end{itemize}
Pauli noise can be composed with any of these three errors to yield a general non-Pauli noise model for the control-$Z$ gates. 


\end{document}